\documentclass[prl,twocolumn,preprintnumbers,superscriptaddress,amsmath,amssymb]{revtex4-1}
\usepackage{graphicx}
\usepackage{subfigure}
\usepackage{mathrsfs}
\usepackage{amsfonts}
\usepackage{times}
\usepackage{amsmath}
\usepackage{leftidx}
\usepackage{tikz}
\usepackage{tikz-network}
\usepackage{color}
\usepackage[colorlinks,linkcolor=blue,citecolor=blue]{hyperref}

\newtheorem{theorem}{Theorem}
\newtheorem{lemma}{Lemma}
\newtheorem{proposition}{Proposition}
\usepackage{bbold}
\usepackage{braket}
\usepackage{mathtools}
\usepackage[normalem]{ulem}

\begin{document}
\title{Long-Range Free Fermions: Lieb-Robinson Bound, Clustering Properties, and Topological Phases}
\author{Zongping Gong}
\affiliation{Max-Planck-Institut f\"ur Quantenoptik, Hans-Kopfermann-Stra{\ss}e 1, D-85748 Garching, Germany}
\affiliation{Munich Center for Quantum Science and Technology, Schellingstra{\ss}e 4, 80799 M\"unchen, Germany}
\author{Tommaso Guaita}
\affiliation{Max-Planck-Institut f\"ur Quantenoptik, Hans-Kopfermann-Stra{\ss}e 1, D-85748 Garching, Germany}
\affiliation{Munich Center for Quantum Science and Technology, Schellingstra{\ss}e 4, 80799 M\"unchen, Germany}
\affiliation{Dahlem Center for Complex Quantum Systems, Freie Universit\"at Berlin, 14195 Berlin, Germany}
\author{J. Ignacio Cirac}
\affiliation{Max-Planck-Institut f\"ur Quantenoptik, Hans-Kopfermann-Stra{\ss}e 1, D-85748 Garching, Germany}
\affiliation{Munich Center for Quantum Science and Technology, Schellingstra{\ss}e 4, 80799 M\"unchen, Germany}
\date{\today}

\begin{abstract}
We consider free fermions living on lattices in arbitrary dimensions, where hopping amplitudes follow a power-law decay with respect to the distance. We focus on the regime where this power is larger than the spatial dimension (i.e., where the single particle energies are guaranteed to be bounded) for which we provide a comprehensive series of fundamental constraints on their equilibrium and nonequilibrium properties. First we derive a Lieb-Robinson bound which is optimal in the spatial tail. This bound then implies a clustering property with essentially the same power law for the Green's function, whenever its variable lies outside the energy spectrum. The widely believed (but yet unproven in this regime) clustering property for the ground-state correlation function follows as a corollary among other implications. Finally, we discuss the impact of these results on topological phases in long-range free-fermion systems: they justify the equivalence between Hamiltonian and state-based definitions and the extension of the short-range phase classification to systems with decay power larger than the spatial dimension. Additionally, we argue that all the short-range topological phases are unified whenever this power is allowed to be smaller.
\end{abstract}
\maketitle

\emph{Introduction.---}Locality is a central concept in quantum many-body physics \cite{Hastings2010}. One of the most important implications of locality, which explicitly means the Hamiltonian is a sum of local terms, is the Lieb-Robinson bound that claims a ``soft" light cone for correlation propagation \cite{Lieb1972,Nachtergaele2006,Bravyi2006}. Further assuming an energy gap in the Hamiltonian, locality implies that the ground-state correlation functions should decay exponentially \cite{Hastings2004a,Hastings2006}. This so-called clustering property gives a partial justification for studying phases of quantum matter \cite{Zeng2019} by focusing on short-range correlated many-body states, which typically obey entanglement area laws \cite{Hastings2007,Eisert2010} and admit efficient representations based on tensor networks \cite{Cirac2021}. 

The past couple of years has witnessed a series of breakthroughs on generalizing the above locality-related results to those ``not-so-local" quantum many-body systems with power-law decaying interactions \cite{Gong2014,FossFeig2015,FossFeig2017,Matsuta2017,Chen2019,Else2020,Kuwahara2020,Tran2020,Kuwahara2021,Tran2021,Wang2022}, commonly dubbed long-range systems \cite{Defenu2021}. This topic is of both fundamental and pratical importance as long-range interactions appear ubiquitously in nature and quantum simulators \cite{Esslinger2013,Tudela2015,Cirac2019,Browaeys2020,Monroe2021}. In particular, the problem of finding Lieb-Robinson bounds with optimal light-cone behaviors has recently been solved for both interacting \cite{Tran2021} and noninteracting (free-fermion) \cite{Tran2020} long-range systems. In contrast, other results such as clustering properties and phase classifications remain to be improved or explored, even on the noninteracting level \cite{Gong2022}. We note that, despite their simplicity, free fermions can already accommodate various topological phases \cite{Ryu2008,Kitaev2009,Ryu2016}, whose long-range generalizations have been considered in various specific models \cite{Vodola2014,Vodola2015,Patrick2017,Lepori2018,Viyuela2018,Francica2022} and may be realized effectively in spin systems such as atomic arrays \cite{Bettles2017,Perczel2017} and NV centers \cite{Kucsko2018}. Also, long-range models appear naturally in the context of fermionic Gaussian projected entangled pair states \cite{Wahl2013,Wahl2014,Dubail2015}.

In this work, we report some essential progress on long-range free fermions, focusing on universal and rigorous results both in and out of equilibrium. First, we derive a new Lieb-Robinson bound as the noninteracting counterpart of that in Ref.~\cite{FossFeig2015}, which is optimal in the spatial tail but not in the light cone. This bound implies an (almost) optimal clustering property for Green's functions, leading to a widely believed ground-state clustering property among other applications. The latter result justifies the equivalence between state and Hamiltonian formalisms for long-range free-fermion topological phases. In addition, we argue that the topological classification of short-range phases remains applicable to long-range phases if the decay power is larger than the spatial dimension, and collapses otherwise.

\emph{Setup.---}For simplicity, we focus on free fermions living on a $d$-dimensional hypercubic lattice $\Lambda\subset\mathbb{Z}^d$ with particle number conservation, where possible internal states (e.g., spin) per site form a set $I$. The generalization to the cases without number conservation and other lattices is straightforward \cite{SM}. Denoting $\hat c^\dag_{\boldsymbol{r}s}$/$\hat c_{\boldsymbol{r}s}$ as the creation/annihilation operator of a fermion with internal state $s\in I$ at site  $\boldsymbol{r}\in\Lambda$, we know that the Hamiltonian generally reads
\begin{equation}
\hat H = \sum_{\substack{\boldsymbol{r},\boldsymbol{r}'\in\Lambda \\ s,s'\in I}} H_{\boldsymbol{r}s, \boldsymbol{r}'s'}\:\hat c^\dag_{\boldsymbol{r}s}\hat c_{\boldsymbol{r}'s'},
\end{equation}
where $H$ is a $|\Lambda||I|\times|\Lambda||I|$ Hermitian matrix. We assume the hopping amplitudes follow a power-law decay. This means for any $|I|\times |I|$ block $[H_{\boldsymbol{r}\boldsymbol{r}'}]_{ss'}\equiv H_{\boldsymbol{r}s,\boldsymbol{r}'s'}$, or equivalently $H_{\boldsymbol{r}\boldsymbol{r}'}\equiv P_{\boldsymbol{r}}HP_{\boldsymbol{r}'}$ with $P_{\boldsymbol{r}}$ being the projector onto site $\boldsymbol{r}$, there exist two positive $\mathcal{O}(1)$ constants \cite{O1} $J$ and $\alpha$ such that its operator norm satisfies
\begin{equation}
\|H_{\boldsymbol{r}\boldsymbol{r}'}\|\le \frac{J}{(|\boldsymbol{r}-\boldsymbol{r}'|+1)^\alpha},
\label{ad}
\end{equation}
where $|\boldsymbol{r}-\boldsymbol{r}'|$ is the distance between $\boldsymbol{r}$ and $\boldsymbol{r}'$. 
We call such a long-range system satisfying Eq.~(\ref{ad}) \emph{$\alpha$-decaying} to highlight the explicit exponent. 

Two comments are in order. First, given a fixed $\alpha$, one can equivalently replace $|\boldsymbol{r}-\boldsymbol{r}'|+1$ by $|\boldsymbol{r}-\boldsymbol{r}'|$ in Eq.~(\ref{ad}) with 
$\boldsymbol{r}\neq \boldsymbol{r}'$ specified. While both are commonly used conventions, 
we prefer Eq.~(\ref{ad}) as we need not 
exclude $\boldsymbol{r}=\boldsymbol{r}'$. Second, one can check that $\alpha > d$ is necessary and sufficient for \emph{any} $\alpha$-decaying Hamiltonian to have bounded single particle energies, i.e., $\|H\|<\infty$, so that the total energy is extensive. Our results are mostly obtained in this thermodynamically stable regime \cite{Kuwahara2021}.

\emph{Lieb-Robinson bound.---}For free fermions 
it suffices to consider individual single particles. Our first main result concerns how fast an initially localized particle propagates under the time evolution governed by $\hat H$: 
\begin{theorem}[Lieb-Robinson bound]
For any $\alpha$-decaying Hamiltonian $H$ with $\alpha > d$, there exists an $\mathcal{O}(1)$ constant $t_{\rm c}$ depending only on $\alpha$ and $d$ such that for any $t>\textcolor{black}{t_{\rm c}}$ 
\begin{equation}
\|P_{\boldsymbol{r}} e^{-iHt} P_{\boldsymbol{r}'}\| \le \frac{K(t)}{(|\boldsymbol{r} - \boldsymbol{r}'| + 1)^\alpha},
\label{LR}
\end{equation}
where $K(t)$ grows polynomially fast in time and $K(t)\propto t^{\alpha (\alpha+1)/(\alpha -d)}$ for large $t$.
\label{Thm:LR}
\end{theorem}
Equation~(\ref{LR}) essentially gives an upper bound on the wavefunction amplitude on site $\boldsymbol{r}$ at time $t$ of a single-particle state initially localized at $\boldsymbol{r}'$. It thus constrains the spreading of wave function in this ``continuous-time quantum walk" setting \cite{Greiner2015b}.

\begin{figure}[!t]
\begin{center}
\includegraphics[width=7cm, clip]{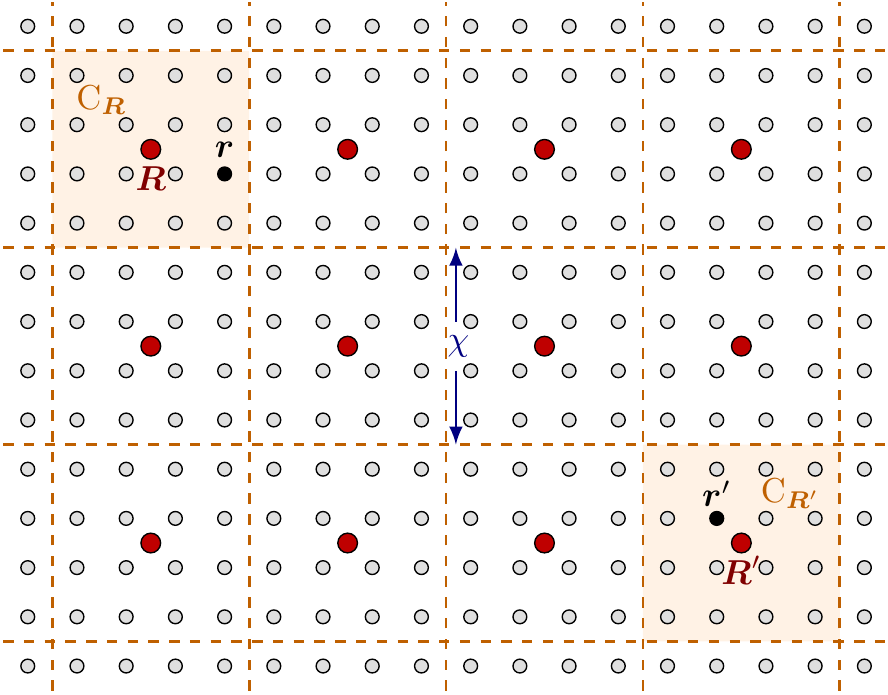}
\caption{Coarse-graining of a square lattice $\Lambda$ (grey circles) into $\tilde \Lambda$ (red circles) with a rescaling $\chi$. Here ${\rm C}_{\boldsymbol{R}}$ denotes all the sites in $\Lambda$, including $\boldsymbol{r}$, that are coarse-grained into $\boldsymbol{R}\in\tilde\Lambda$. The coarse-grained projector is thus defined as $P_{\boldsymbol{R}}\equiv\sum_{\boldsymbol{r}\in{\rm C}_{\boldsymbol{R}}}P_{\boldsymbol{r}}$. Note that no periodicity of the Hamiltonian is assumed.}
      \label{SL}
      \end{center}
\end{figure}

This bound (\ref{LR}) appears to be rather similar to the bound in Ref.~\cite{FossFeig2015} for interacting long-range systems, but a crucial difference here is that in our (free) case it holds for the whole $\alpha > d$ regime while the interacting case requires $\alpha > 2d$, as we will explain for the derivation in the next paragraph. As is also the case in Ref.~\cite{FossFeig2015}, the time scaling of $K(t)$ in Eq.~(\ref{LR}) is far from optimal. Indeed, the light cone $t\propto|\boldsymbol{r}-\boldsymbol{r}'|^{(\alpha-d)/(\alpha+1)}$ is linear only in the short-range limit $\alpha\to\infty$, while optimally it would be linear already for $\alpha>d+1$ \cite{Tran2020}. On the other hand, the spatial tail of Eq.~(\ref{LR}) is optimal. To see this, we only have to consider $\hat H= J/(|\boldsymbol{r}-\boldsymbol{r}'|+1)^\alpha(\hat c^\dag_{\boldsymbol{r}} \hat c_{\boldsymbol{r}'} + {\rm H.c.})$. Then we have $\|P_{\boldsymbol{r}} e^{-iHt} P_{\boldsymbol{r}'}\|\ge 2t/[\pi(|\boldsymbol{r}-\boldsymbol{r}'|+1)^\alpha]$ at large distance, i.e. $\pi(|\boldsymbol{r} - \boldsymbol{r}'|+1)^\alpha/(2J)>t$. It is also worthwhile to compare this bound (\ref{LR}) to the free-fermion bound in Ref.~\cite{Tran2020}, which is optimal in the light cone but not in the tail. 

Let us outline the proof of Theorem~\ref{Thm:LR}. It is instructive to first recall that a direct Taylor expansion of $e^{-iHt}$ gives a bound like Eq.~(\ref{LR}) but with $K(t)\propto e^{\lambda t}$ \cite{Hastings2006}. To tighten this exponential dependence, the basic idea is to separate the Hamiltonian into the short-range and long-range parts, i.e., $H=H_{\rm sr}+H_{\rm lr}$, where the short-range part is determined $[H_{\rm sr}]_{\boldsymbol{r}s,\boldsymbol{r}'s'}= H_{\boldsymbol{r}s,\boldsymbol{r}'s'}$ for $|\boldsymbol{r}-\boldsymbol{r}'|\le \chi$ ($\chi$: cutoff parameter) but otherwise $[H_{\rm sr}]_{\boldsymbol{r}s,\boldsymbol{r}'s'}=0$. We can then work in the interaction picture with respect to the former: 
\begin{equation}
    e^{-iHt}= e^{-iH_{\rm sr}t}{\rm T}e^{-i\int^t_{0} dt' H^{(\rm I)}_{\rm lr}(t')},
    \label{ip}
\end{equation}
where ${\rm T}$ denotes the time ordering and $H^{(\rm I)}_{\rm lr}(t)=e^{iH_{\rm sr}t}H_{\rm lr}e^{-iH_{\rm sr}t}$. By Taylor expansion in the interaction picture, we can also obtain an exponential factor but with a modified coefficient $\lambda_\chi$, which can be made sufficiently small by properly choosing $\chi$. While so far the procedure largely follows Ref.~\cite{FossFeig2015}, a crucial difference here is that we further perform a \emph{coarse graining} of the lattice $\Lambda$ into $\tilde\Lambda$ at the same scale $\chi$ (see Fig.~\ref{SL}). This helps us get rid of a factor $\chi^d$ in $\lambda_\chi$ compared to the interacting case, making it proportional to $\chi^{-(\alpha-d)}$ rather than $\chi^{-(\alpha-2d)}$. Therefore, $\alpha>d$ is enough for suppressing $\lambda_\chi t$ by choosing a sufficiently large $\chi$.

To further illustrate how and why the coarse graining works, we first write down the Taylor-expansion bound on the left-hand side of Eq.~(\ref{LR}) in the interaction picture (\ref{ip}) \cite{PA}: 
\begin{equation}
\begin{split}
&\| P_{\boldsymbol{r}} e^{-iHt} P_{\boldsymbol{r}'}\| \le \sum^\infty_{n=0} \int^t_{0} dt_n\int^{t_n}_0 dt_{n-1} \cdots \int^{t_2}_0 dt_1 \\
\times&\left\| P_{\boldsymbol{r}} e^{-iH_{\rm sr}(t-t_n)}\overleftarrow{\prod}^n_{m=1} H_{\rm lr} e^{-iH_{\rm sr}(t_m-t_{m-1})}
P_{\boldsymbol{r}'}\right\|,
\end{split}
\label{ipte}
\end{equation}
where $t_0\equiv0$. Instead of inserting $\mathbb{1}=\sum_{\boldsymbol{r}\in\Lambda}P_{\boldsymbol{r}}$ ($\mathbb{1}$: identity) as is essentially the strategy used in Ref.~\cite{FossFeig2015}, we insert the coarse-grained decomposition $\mathbb{1}=\sum_{\boldsymbol{R}\in\tilde\Lambda}P_{\boldsymbol{R}}$ (see Fig.~\ref{SL}) so that each integrand in Eq.~(\ref{ipte}) can be upper bounded by
\begin{widetext}
\begin{equation}
    \sum_{\{\boldsymbol{R}_j\in\tilde\Lambda\}^{2n}_{j=1}}  \|P_{\boldsymbol{R}} e^{-iH_{\rm sr}(t-t_n)} P_{\boldsymbol{R}_{2n}}\|
    \prod^n_{m=1}\|P_{\boldsymbol{R}_{2m}}H_{\rm lr}P_{\boldsymbol{R}_{2m-1}} \| \|P_{\boldsymbol{R}_{2m-1}} e^{-iH_{\rm sr}(t_m-t_{m-1})}P_{\boldsymbol{R}_{2m-2}}\|,
\label{PRB}
\end{equation}
\end{widetext}
where $\boldsymbol{R}_0\equiv \boldsymbol{R}'$ and $\boldsymbol{R},\boldsymbol{R}'$ are determined such that they include $\boldsymbol{r},\boldsymbol{r}'$ respectively.
Obviously, except for the two boundary factors, the bulk product is always smaller 
than the refined decomposition (to each lattice site). In fact it turns out to be smaller by a factor $\chi^{-nd}$ which leads to the qualitative improvement of $\lambda_\chi$ discussed above. This is because each $\|P_{\boldsymbol{R}_{2m}}H_{\rm lr}P_{\boldsymbol{R}_{2m-1}}\|$ is roughly smaller than the corresponding sum of $
\|P_{\boldsymbol{r}_{2m}}H_{\rm lr}P_{\boldsymbol{r}_{2m-1}}\|$ by a factor $\chi^d$ ($\boldsymbol{r}_m\in\Lambda$ is a site coarse-grained into $\boldsymbol{R}_m\in\tilde\Lambda$), while $\|P_{\boldsymbol{R}_{2m-1}} e^{-iH_{\rm sr}(t_m-t_{m-1})}P_{\boldsymbol{R}_{2m-2}}\|$ differs from $\|P_{\boldsymbol{r}_{2m-1}} e^{-iH_{\rm sr}(t_m-t_{m-1})}P_{\boldsymbol{r}_{2m-2}}\|$ by mostly an $\mathcal{O}(1)$ factor \cite{SM}. Note that in the interacting case this improvement is canceled by a factor of $\chi^d$ from (the interacting counterpart of) each $\|P_{\boldsymbol{R}_{2m-1}} e^{-iH_{\rm sr}(t_m-t_{m-1})}P_{\boldsymbol{R}_{2m-2}}\|$, accounting for the size of support of $P_{\boldsymbol{R}}$. It is the single-particle nature of free systems that allows us not to ``pay the price".

\emph{Clustering properties.---}We move on to introduce the second main result 
--- the clustering property of the Green's function (or resolvent \cite{Kato1966})
\begin{equation}
G(z)\equiv(z-H)^{-1}.
\label{GF}
\end{equation}
We assume $z$ is outside the spectrum of $H$ and we define $\Delta(z)\equiv\|G(z)\|^{-1}$ as the distance of $z$ to such spectrum.
Precisely speaking, we have:
\begin{theorem}[Clustering property of the Green's function]
For an $\alpha$-decaying Hamiltonian $H$ with $\alpha>d$ \textcolor{black}{and 
$z\in\mathbb{C}$ that} is not an eigenvalue of $H$, 
the Green's function \eqref{GF} satisfies
\begin{equation}
\|G_{\boldsymbol{r}\boldsymbol{r}'}(z)\|\le \frac{{\rm poly}(\log(|\boldsymbol{r}-\boldsymbol{r}'|+1))}{(|\boldsymbol{r}-\boldsymbol{r}'|+1)^\alpha},
\label{CPGF}
\end{equation}
where $G_{\boldsymbol{r}\boldsymbol{r}'}(z)=P_{\boldsymbol{r}}G(z)P_{\boldsymbol{r}'}$ and $\rm poly(\cdot)$ means a polynomially large function with $|\boldsymbol{r}-\boldsymbol{r}'|$-independent coefficients, which nevertheless depend on $\Delta(z)$ and diverge for $\Delta(z)\to0$.  
\label{Thm:GF}
\end{theorem}
Here the condition $\Delta(z)\neq0$ is absolutely necessary since otherwise even short-range hopping ($\alpha\to\infty$) can generate long-range correlations and interactions, manifesting as, for instance, Friedel oscillations \cite{Cheianov2006} and RKKY interactions \cite{Brey2007} in the presence of impurities. The short-range counterpart of Theorem~\ref{Thm:GF} has been considered in Ref.~\cite{Watanabe2018}. 


A direct corollary of Theorem~\ref{Thm:GF} is the clustering property of ground-state correlation functions. In the case of free fermions, it is natural to consider the covariance matrix: 
\begin{equation}
    C_{\boldsymbol{r}s,\boldsymbol{r}'s'}\equiv\langle\Psi_0|\hat c_{\boldsymbol{r}'s'}^\dag \hat c_{\boldsymbol{r}s}|\Psi_0\rangle,
    \label{cm}
\end{equation} 
where $|\Psi_0\rangle$ is the 
ground state of $\hat H$. Without loss of generality, we may assume the Fermi energy, which lies in a band gap, to be zero, so that \cite{Vishwanath2010,Hughes2011}
\begin{equation}
2C=\mathbb{1} - {\rm sgn} H.
\label{CH}
\end{equation}
Thanks to Wick's theorem, any correlation functions can be obtained from the covariance matrix (\ref{cm}), so it suffices to consider the clustering properties for the latter, i.e., a bound on $\|C_{\boldsymbol{r}\boldsymbol{r}'}\|$ with $C_{\boldsymbol{r}\boldsymbol{r}'}\equiv P_{\boldsymbol{r}}CP_{\boldsymbol{r}'}$. Note that  
\begin{equation}
C= 
\oint_{\ell_<} \frac{dz}{2\pi i} G(z),
\label{CG}
\end{equation}
where $\ell_<$ is a closed loop that encompasses all the bands on the negative real axis, i.e., below the Fermi energy. Since the length of $\ell_<$ is bounded by a constant (due to the finiteness of $\|H\|$) while $\|G_{\boldsymbol{r}\boldsymbol{r}'}(z)\|$ satisfies Eq.~(\ref{CPGF}) $\forall z\in \ell_<$, we know that $\|C_{\boldsymbol{r}\boldsymbol{r}'}\|$ also satisfies Eq.~(\ref{CPGF}). Moreover, Theorem~\ref{Thm:GF} has broader implications. For example, it implies that any bound state outside the spectrum induced by an impurity supported on $\mathcal{O}(1)$ sites has an algebraically decaying profile in real space, with essentially the same exponent $\alpha$. This can be seen from $\boldsymbol{\psi}_{\rm b} = G(E_{\rm b})V \boldsymbol{\psi}_{\rm b}$, where $\boldsymbol{\psi}_{\rm b}$ is the wave function of the bound state with eigenenergy $E_{\rm b}$ and $V$ is the impurity potential \cite{Slager2015}. We will again exploit Theorem~\ref{Thm:GF} when discussing topological phases in the next section.

Finally, let us sketch out the proof of Theorem~\ref{Thm:GF} \cite{SM}. Similar to Refs.~\cite{Hastings2004a,Hastings2004b,Hastings2006,Hernandez2017,Tran2020} concerning ground-state correlations, the main idea is to construct an analytic filter function $f_\sigma(t)$ such that it decays rapidly for large $t$ and its Fourier transform $\mathfrak{F}[f_\sigma](\omega)\equiv \int^\infty_{-\infty}\frac{dt}{2\pi}f_\sigma(t)e^{-i\omega t}$ (also analytic) well approximates $(z-\omega)^{-1}$ if $\omega$ is away from $z$ by a few $\sigma$'s, a control parameter to be determined later. Via Eq.~(\ref{GF}) and up to some error terms involving $\sigma$, such a filter enables us to express 
$G_{\boldsymbol{r}\boldsymbol{r}'}(z)$ as 
\begin{equation}
P_{\boldsymbol{r}}\mathfrak{F}[f_\sigma](H)P_{\boldsymbol{r}'}=\int^\infty_{-\infty}\frac{dt}{2\pi}f_\sigma(t)P_{\boldsymbol{r}}e^{-iHt}P_{\boldsymbol{r}'},
\end{equation}
which can be bounded by the Lieb-Robinson bound (\ref{LR}) (and the trivial bound $\|P_{\boldsymbol{r}}e^{-iHt}P_{\boldsymbol{r}'}\|\le1$ for late times). The desired bound (\ref{CPGF}) is obtained by choosing an appropriate $\sigma$, which turns out to be proportional to $\Delta(z)$ and sub-logarithmically suppressed by $|\boldsymbol{r}-\boldsymbol{r}'|$.


\emph{Topological phases.---}Free-fermion 
topological phases 
are defined in terms of equivalence classes under continuous deformations of gapped quadratic Hamiltonians or alternatively of free-fermion (Gaussian) 
states. In the short-range case these two approaches can be readily seen to be equivalent \cite{Gong2021}. 
In the long-range case, given an $\alpha$-decaying free-fermion 
state, it is easy to construct a parent Hamiltonian that is also $\alpha$-decaying 
\textcolor{black}{by taking} $H=\mathbb{1}-2C$ (cf. Eq.~(\ref{CH})). It follows that a continuous deformation of a state implies that of the parent Hamiltonian.

According to the clustering properties 
proven above, we know that the converse is also true if $\alpha>d$. Given a continuous path of gapped $\alpha$-decaying Hamiltonians $H_\lambda$ parametrized by $\lambda\in[0,1]$, their ground states will also be (almost) $\alpha$-decaying due to the clustering property of ground-state correlations. It can be further shown that they define a continuous path in the space of states by using the clustering property of the Green's function. 
Indeed, we have
\begin{equation}
C_{\lambda'} - C_{\lambda} = \oint_{\ell_<}\frac{dz}{2\pi i} G_\lambda(z)(H_{\lambda'} - H_{\lambda})G_{\lambda'}(z) \,,
\end{equation} 
where $\ell_<$ encircles the lower bands of both $H_\lambda$ and $H_{\lambda'}$, which is always possible given a minimal gap during the deformation. Due to Theorem~\ref{Thm:GF}, 
we have that $\|G_\lambda(z)\|\leq \max_{\boldsymbol{r}}\sum_{\boldsymbol{r}'}\|G_{\lambda,\,\boldsymbol{r}\boldsymbol{r}'}(z)\|$ is bounded along $\ell_<$ , implying that $C_\lambda$ depends continuously on $H_\lambda$.

This analysis justifies the equivalence of considering free-fermion 
states and gapped quadratic Hamiltonians for $\alpha > d$. In this case we can also say something more about the structure of existing phases. One can show that every long-range gapped Hamiltonian with $\alpha>d$ is continuously connected to a short-range one, implying that there are no new phases unique to long-range Hamiltonians. To see this, consider the Hamiltonians defined by $H_{\kappa,\boldsymbol{r}\boldsymbol{r}'} 
=e^{-\kappa |\boldsymbol{r}-\boldsymbol{r'}|}\,H_{\boldsymbol{r}\boldsymbol{r'}}$ which constitute a continuous path with respect to $\kappa$ and 
can be shown to be gapped for sufficiently small but finite $\kappa$ \cite{SM}. This path connects the long-range Hamiltonian at $\kappa=0$ to a short-range one (i.e., exponentially decaying) at finite $\kappa$.

Furthermore, there should be no unification of short-range phases in the regime $\alpha > d$, since for translation-invariant systems the Bloch Hamiltonians 
$h(\boldsymbol{k})$ remain continuous in $\boldsymbol{k}$ and all the short-range topological invariants remain well-defined and robust under continuous deformations of the Hamiltonian. For disordered systems, 
the index theorem of Ref.~\cite{Katsura2018} shows that topological invariants must remain equal to a fixed integer along any path $H_\lambda$ provided that $C$ changes continuously with respect to $H$, which we have shown above to be true.

\begin{figure}
    \centering
    \includegraphics[width=0.5\textwidth]{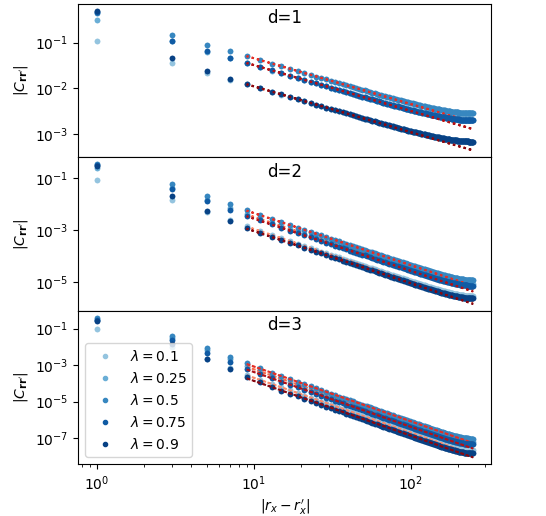}
    \caption{Decay rates of $\|C_{\boldsymbol{r}\boldsymbol{r}'}\|$ for the ground states of the Hamiltonians $h_\lambda$ defined by $h_\lambda=(1-2\lambda)h_0+2\lambda h_{\rm fD}$ for $\lambda\in[0,0.5]$ and $h_\lambda=2(1-\lambda) h_{\rm fD}+(2\lambda-1)h_{\rm Topo}$ for $\lambda\in[0.5,1]$ in dimensions $d=1,2,3$. The red dashed lines are a fit of the long distance behavior of the data with $\|C_{\boldsymbol{r}\boldsymbol{r}'}\|\propto|\boldsymbol{r}-\boldsymbol{r}'|^{-d}$, implying the ill-definedness of conventional topological numbers. The numerical calculations are performed on finite hypercubic lattices of side $L=500$ with anti-periodic boundary conditions \cite{APBC}. In all cases the odd sites along the $x$-axis are plotted, as this is the subset of sites 
    in the lattice that shows the slowest decay. }
    \label{fig:Gamma_l}
\end{figure}

Remarkably, the threshold $\alpha=d$ above which the short-range paradigm persists is optimal, i.e., cannot be improved to be smaller. This is because 
if we allow $\alpha< d$ then all the \emph{short-range} topological phases are expected to be 
unified (up to a 0D topological invariant such as the fermion number parity). Without loss of generality, we focus on translation-invariant representatives described by Bloch Hamiltonians. The argument is based on the well known result that all the short-range topological phases can be obtained by perturbing a Dirac Hamiltonian with a mass term \cite{Ryu2016}
\begin{equation}
h(\boldsymbol{k}) = \sum^d_{\mu=1} \sin k_\mu \Gamma_\mu - \left(\sum^d_{\mu=1}\cos k_\mu - m\right)\Gamma_0,
\end{equation}
where ${\left\{\Gamma_\mu\right\}}_{\mu=0,\dots,d}$ are Hermitian Dirac matrices satisfying $\{\Gamma_\mu,\Gamma_\nu\}=2\delta_{\mu\nu}$. Decreasing $m$ from $m>d$ to $m\in(d-2,d)$, there is a single band crossing at $\boldsymbol{k}=\boldsymbol{0}$, giving rise to a transition from a trivial phase to a topological phase with unit topological number \cite{Lee2019}. Let us take, for instance, the topological Bloch Hamiltonian $h_{\mathrm{Topo}}(\boldsymbol{k})$ defined by choosing $m=d-1$. We can show that $h_{\mathrm{Topo}}(\boldsymbol{k})$ is connected to the trivial Hamiltonian $h_0(\boldsymbol{k})=\Gamma_0$ through a continuous path of long-range gapped Hamiltonians, provided that $\alpha$ is not constrained to be larger than $d$. To this end, we consider a linear interpolation from $h_0(\boldsymbol{k})$ to $h_{\rm fD}(\boldsymbol{k})$ and then from $h_{\rm fD}(\boldsymbol{k})$ to $h_{\mathrm{Topo}}(\boldsymbol{k})$, where $h_{\rm fD}(\boldsymbol{k})$ is the flattened Dirac Hamiltonian 
\begin{equation}
h_{\rm fD}(\boldsymbol{k})= \sum^d_{\mu=1} \frac{\sin k_\mu}{\sqrt{\sum^d_{\mu=1}\sin^2k_\mu}}\Gamma_\mu \,.
\end{equation}
One can see that $h(\boldsymbol{k})^2>0$ during the whole deformation, meaning that the gap does not close. Furthermore $h_{\rm fD}(\boldsymbol{k})$ is (almost) $d$-decaying in real space~\cite{SM}, while $h_0(\boldsymbol{k})$ and $h_{\mathrm{Topo}}(\boldsymbol{k})$ are local, so the Hamiltonian is at most as nonlocal as $d$-decaying during the deformation. Numerical analysis suggests that also the ground state covariance matrix $C$ remains always $d$-decaying along such path, as shown in Fig.~\ref{fig:Gamma_l}.

\emph{Summary and outlook.---}We have derived a tight (in the sense of spatial tail) Lieb-Robinson bound for $\alpha$-decaying free-fermion systems with $\alpha>d$. This bound allows us to prove an (almost) optimal clustering property for the Green's function, which implies the clustering property for the ground-state correlations in gapped systems. These results justify the equivalence between state and Hamiltonian-based definitions of topological phases in long-range free-fermion systems. In addition, we argue that all the short-range topological phases are connected within the space of $\alpha$-decaying systems with $\alpha<d$.

A relevant open problem is how to further improve the Lieb-Robinson bound to be consistent with the optimal light cone \cite{Tran2020}. Also, one still has to examine the validity of bulk-edge correspondence \cite{Jones2022} and prove the clustering properties for topological edge modes localized at sharp edges, where our local-impurity argument does not apply. Improving the entanglement area law \cite{FossFeig2017,Kuwahara2020a} for long-range free fermions could be yet another direction of future study. 
One may also consider whether our progress can facilitate the long-range generalization of the clustering properties for short-range Anderson localized systems \cite{Frohlich1983,Aizenman1993}. 

Z.G. is supported by the Max-Planck-Harvard Research Center for Quantum Optics (MPHQ). T.G. is supported by the Deutsche Forschungs- gemeinschaft (DFG, German Research Foundation) under Germany’s Excellence Strategy – EXC-2111 – 39081486. J.I.C. acknowledges support by the EU Horizon 2020 program through the ERC Advanced Grant QENOCOBA No. 742102. 

\bibliography{GZP_references} 

\begin{thebibliography}{66}%
\makeatletter
\providecommand \@ifxundefined [1]{%
 \@ifx{#1\undefined}
}%
\providecommand \@ifnum [1]{%
 \ifnum #1\expandafter \@firstoftwo
 \else \expandafter \@secondoftwo
 \fi
}%
\providecommand \@ifx [1]{%
 \ifx #1\expandafter \@firstoftwo
 \else \expandafter \@secondoftwo
 \fi
}%
\providecommand \natexlab [1]{#1}%
\providecommand \enquote  [1]{``#1''}%
\providecommand \bibnamefont  [1]{#1}%
\providecommand \bibfnamefont [1]{#1}%
\providecommand \citenamefont [1]{#1}%
\providecommand \href@noop [0]{\@secondoftwo}%
\providecommand \href [0]{\begingroup \@sanitize@url \@href}%
\providecommand \@href[1]{\@@startlink{#1}\@@href}%
\providecommand \@@href[1]{\endgroup#1\@@endlink}%
\providecommand \@sanitize@url [0]{\catcode `\\12\catcode `\$12\catcode
  `\&12\catcode `\#12\catcode `\^12\catcode `\_12\catcode `\%12\relax}%
\providecommand \@@startlink[1]{}%
\providecommand \@@endlink[0]{}%
\providecommand \url  [0]{\begingroup\@sanitize@url \@url }%
\providecommand \@url [1]{\endgroup\@href {#1}{\urlprefix }}%
\providecommand \urlprefix  [0]{URL }%
\providecommand \Eprint [0]{\href }%
\providecommand \doibase [0]{http://dx.doi.org/}%
\providecommand \selectlanguage [0]{\@gobble}%
\providecommand \bibinfo  [0]{\@secondoftwo}%
\providecommand \bibfield  [0]{\@secondoftwo}%
\providecommand \translation [1]{[#1]}%
\providecommand \BibitemOpen [0]{}%
\providecommand \bibitemStop [0]{}%
\providecommand \bibitemNoStop [0]{.\EOS\space}%
\providecommand \EOS [0]{\spacefactor3000\relax}%
\providecommand \BibitemShut  [1]{\csname bibitem#1\endcsname}%
\let\auto@bib@innerbib\@empty
\bibitem [{\citenamefont {Hastings}(2010)}]{Hastings2010}%
  \BibitemOpen
  \bibfield  {author} {\bibinfo {author} {\bibfnamefont {M.~B.}\ \bibnamefont
  {Hastings}},\ }\href@noop {} {\enquote {\bibinfo {title} {Locality in quantum
  systems},}\ } (\bibinfo {year} {2010}),\ \bibinfo {note}
  {arXiv:1008.5137}\BibitemShut {NoStop}%
\bibitem [{\citenamefont {Lieb}\ and\ \citenamefont
  {Robinson}(1972)}]{Lieb1972}%
  \BibitemOpen
  \bibfield  {author} {\bibinfo {author} {\bibfnamefont {E.~H.}\ \bibnamefont
  {Lieb}}\ and\ \bibinfo {author} {\bibfnamefont {D.~W.}\ \bibnamefont
  {Robinson}},\ }\href {https://doi.org/10.1007/978-3-662-10018-9_25}
  {\bibfield  {journal} {\bibinfo  {journal} {Commun. Math. Phys.}\ }\textbf
  {\bibinfo {volume} {28}},\ \bibinfo {pages} {251} (\bibinfo {year}
  {1972})}\BibitemShut {NoStop}%
\bibitem [{\citenamefont {Nachtergaele}\ and\ \citenamefont
  {Sims}(2006)}]{Nachtergaele2006}%
  \BibitemOpen
  \bibfield  {author} {\bibinfo {author} {\bibfnamefont {B.}~\bibnamefont
  {Nachtergaele}}\ and\ \bibinfo {author} {\bibfnamefont {R.}~\bibnamefont
  {Sims}},\ }\href {\doibase 10.1007/s00220-006-1556-1} {\bibfield  {journal}
  {\bibinfo  {journal} {Commun. Math. Phys.}\ }\textbf {\bibinfo {volume}
  {265}},\ \bibinfo {pages} {119} (\bibinfo {year} {2006})}\BibitemShut
  {NoStop}%
\bibitem [{\citenamefont {Bravyi}\ \emph {et~al.}(2006)\citenamefont {Bravyi},
  \citenamefont {Hastings},\ and\ \citenamefont {Verstraete}}]{Bravyi2006}%
  \BibitemOpen
  \bibfield  {author} {\bibinfo {author} {\bibfnamefont {S.}~\bibnamefont
  {Bravyi}}, \bibinfo {author} {\bibfnamefont {M.~B.}\ \bibnamefont
  {Hastings}}, \ and\ \bibinfo {author} {\bibfnamefont {F.}~\bibnamefont
  {Verstraete}},\ }\href {\doibase 10.1103/PhysRevLett.97.050401} {\bibfield
  {journal} {\bibinfo  {journal} {Phys. Rev. Lett.}\ }\textbf {\bibinfo
  {volume} {97}},\ \bibinfo {pages} {050401} (\bibinfo {year}
  {2006})}\BibitemShut {NoStop}%
\bibitem [{\citenamefont {Hastings}(2004{\natexlab{a}})}]{Hastings2004a}%
  \BibitemOpen
  \bibfield  {author} {\bibinfo {author} {\bibfnamefont {M.~B.}\ \bibnamefont
  {Hastings}},\ }\href {\doibase 10.1103/PhysRevLett.93.140402} {\bibfield
  {journal} {\bibinfo  {journal} {Phys. Rev. Lett.}\ }\textbf {\bibinfo
  {volume} {93}},\ \bibinfo {pages} {140402} (\bibinfo {year}
  {2004}{\natexlab{a}})}\BibitemShut {NoStop}%
\bibitem [{\citenamefont {Hastings}\ and\ \citenamefont
  {Koma}(2006)}]{Hastings2006}%
  \BibitemOpen
  \bibfield  {author} {\bibinfo {author} {\bibfnamefont {M.~B.}\ \bibnamefont
  {Hastings}}\ and\ \bibinfo {author} {\bibfnamefont {T.}~\bibnamefont
  {Koma}},\ }\href {\doibase 10.1007/s00220-006-0030-4} {\bibfield  {journal}
  {\bibinfo  {journal} {Comm. Math. Phys.}\ }\textbf {\bibinfo {volume}
  {265}},\ \bibinfo {pages} {781} (\bibinfo {year} {2006})}\BibitemShut
  {NoStop}%
\bibitem [{\citenamefont {Zeng}\ \emph {et~al.}(2019)\citenamefont {Zeng},
  \citenamefont {Chen}, \citenamefont {Zhou},\ and\ \citenamefont
  {Wen}}]{Zeng2019}%
  \BibitemOpen
  \bibfield  {author} {\bibinfo {author} {\bibfnamefont {B.}~\bibnamefont
  {Zeng}}, \bibinfo {author} {\bibfnamefont {X.}~\bibnamefont {Chen}}, \bibinfo
  {author} {\bibfnamefont {D.-L.}\ \bibnamefont {Zhou}}, \ and\ \bibinfo
  {author} {\bibfnamefont {X.-G.}\ \bibnamefont {Wen}},\ }\href@noop {} {\emph
  {\bibinfo {title} {Quantum information meets quantum matter}}}\ (\bibinfo
  {publisher} {Springer, New York},\ \bibinfo {year} {2019})\BibitemShut
  {NoStop}%
\bibitem [{\citenamefont {Hastings}(2007)}]{Hastings2007}%
  \BibitemOpen
  \bibfield  {author} {\bibinfo {author} {\bibfnamefont {M.~B.}\ \bibnamefont
  {Hastings}},\ }\href {\doibase 10.1088/1742-5468/2007/08/P08024} {\bibfield
  {journal} {\bibinfo  {journal} {J. Stat. Mech.}\ ,\ \bibinfo {pages}
  {P08024}} (\bibinfo {year} {2007})}\BibitemShut {NoStop}%
\bibitem [{\citenamefont {Eisert}\ \emph {et~al.}(2010)\citenamefont {Eisert},
  \citenamefont {Cramer},\ and\ \citenamefont {Plenio}}]{Eisert2010}%
  \BibitemOpen
  \bibfield  {author} {\bibinfo {author} {\bibfnamefont {J.}~\bibnamefont
  {Eisert}}, \bibinfo {author} {\bibfnamefont {M.}~\bibnamefont {Cramer}}, \
  and\ \bibinfo {author} {\bibfnamefont {M.~B.}\ \bibnamefont {Plenio}},\
  }\href {\doibase 10.1103/RevModPhys.82.277} {\bibfield  {journal} {\bibinfo
  {journal} {Rev. Mod. Phys.}\ }\textbf {\bibinfo {volume} {82}},\ \bibinfo
  {pages} {277} (\bibinfo {year} {2010})}\BibitemShut {NoStop}%
\bibitem [{\citenamefont {Cirac}\ \emph {et~al.}(2021)\citenamefont {Cirac},
  \citenamefont {P\'erez-Garc\'{\i}a}, \citenamefont {Schuch},\ and\
  \citenamefont {Verstraete}}]{Cirac2021}%
  \BibitemOpen
  \bibfield  {author} {\bibinfo {author} {\bibfnamefont {J.~I.}\ \bibnamefont
  {Cirac}}, \bibinfo {author} {\bibfnamefont {D.}~\bibnamefont
  {P\'erez-Garc\'{\i}a}}, \bibinfo {author} {\bibfnamefont {N.}~\bibnamefont
  {Schuch}}, \ and\ \bibinfo {author} {\bibfnamefont {F.}~\bibnamefont
  {Verstraete}},\ }\href {\doibase 10.1103/RevModPhys.93.045003} {\bibfield
  {journal} {\bibinfo  {journal} {Rev. Mod. Phys.}\ }\textbf {\bibinfo {volume}
  {93}},\ \bibinfo {pages} {045003} (\bibinfo {year} {2021})}\BibitemShut
  {NoStop}%
\bibitem [{\citenamefont {Gong}\ \emph {et~al.}(2014)\citenamefont {Gong},
  \citenamefont {Foss-Feig}, \citenamefont {Michalakis},\ and\ \citenamefont
  {Gorshkov}}]{Gong2014}%
  \BibitemOpen
  \bibfield  {author} {\bibinfo {author} {\bibfnamefont {Z.-X.}\ \bibnamefont
  {Gong}}, \bibinfo {author} {\bibfnamefont {M.}~\bibnamefont {Foss-Feig}},
  \bibinfo {author} {\bibfnamefont {S.}~\bibnamefont {Michalakis}}, \ and\
  \bibinfo {author} {\bibfnamefont {A.~V.}\ \bibnamefont {Gorshkov}},\ }\href
  {\doibase 10.1103/PhysRevLett.113.030602} {\bibfield  {journal} {\bibinfo
  {journal} {Phys. Rev. Lett.}\ }\textbf {\bibinfo {volume} {113}},\ \bibinfo
  {pages} {030602} (\bibinfo {year} {2014})}\BibitemShut {NoStop}%
\bibitem [{\citenamefont {Foss-Feig}\ \emph {et~al.}(2015)\citenamefont
  {Foss-Feig}, \citenamefont {Gong}, \citenamefont {Clark},\ and\ \citenamefont
  {Gorshkov}}]{FossFeig2015}%
  \BibitemOpen
  \bibfield  {author} {\bibinfo {author} {\bibfnamefont {M.}~\bibnamefont
  {Foss-Feig}}, \bibinfo {author} {\bibfnamefont {Z.-X.}\ \bibnamefont {Gong}},
  \bibinfo {author} {\bibfnamefont {C.~W.}\ \bibnamefont {Clark}}, \ and\
  \bibinfo {author} {\bibfnamefont {A.~V.}\ \bibnamefont {Gorshkov}},\ }\href
  {\doibase 10.1103/PhysRevLett.114.157201} {\bibfield  {journal} {\bibinfo
  {journal} {Phys. Rev. Lett.}\ }\textbf {\bibinfo {volume} {114}},\ \bibinfo
  {pages} {157201} (\bibinfo {year} {2015})}\BibitemShut {NoStop}%
\bibitem [{\citenamefont {Gong}\ \emph {et~al.}(2017)\citenamefont {Gong},
  \citenamefont {Foss-Feig}, \citenamefont {Brand\~ao},\ and\ \citenamefont
  {Gorshkov}}]{FossFeig2017}%
  \BibitemOpen
  \bibfield  {author} {\bibinfo {author} {\bibfnamefont {Z.-X.}\ \bibnamefont
  {Gong}}, \bibinfo {author} {\bibfnamefont {M.}~\bibnamefont {Foss-Feig}},
  \bibinfo {author} {\bibfnamefont {F.~G. S.~L.}\ \bibnamefont {Brand\~ao}}, \
  and\ \bibinfo {author} {\bibfnamefont {A.~V.}\ \bibnamefont {Gorshkov}},\
  }\href {\doibase 10.1103/PhysRevLett.119.050501} {\bibfield  {journal}
  {\bibinfo  {journal} {Phys. Rev. Lett.}\ }\textbf {\bibinfo {volume} {119}},\
  \bibinfo {pages} {050501} (\bibinfo {year} {2017})}\BibitemShut {NoStop}%
\bibitem [{\citenamefont {Matsuta}\ \emph {et~al.}(2017)\citenamefont
  {Matsuta}, \citenamefont {Koma},\ and\ \citenamefont
  {Nakamura}}]{Matsuta2017}%
  \BibitemOpen
  \bibfield  {author} {\bibinfo {author} {\bibfnamefont {T.}~\bibnamefont
  {Matsuta}}, \bibinfo {author} {\bibfnamefont {T.}~\bibnamefont {Koma}}, \
  and\ \bibinfo {author} {\bibfnamefont {S.}~\bibnamefont {Nakamura}},\ }\href
  {\doibase 10.1007/s00023-016-0526-1} {\bibfield  {journal} {\bibinfo
  {journal} {Ann. Henri Poincar\'e}\ }\textbf {\bibinfo {volume} {18}},\
  \bibinfo {pages} {519} (\bibinfo {year} {2017})}\BibitemShut {NoStop}%
\bibitem [{\citenamefont {Chen}\ and\ \citenamefont {Lucas}(2019)}]{Chen2019}%
  \BibitemOpen
  \bibfield  {author} {\bibinfo {author} {\bibfnamefont {C.-F.}\ \bibnamefont
  {Chen}}\ and\ \bibinfo {author} {\bibfnamefont {A.}~\bibnamefont {Lucas}},\
  }\href {\doibase 10.1103/PhysRevLett.123.250605} {\bibfield  {journal}
  {\bibinfo  {journal} {Phys. Rev. Lett.}\ }\textbf {\bibinfo {volume} {123}},\
  \bibinfo {pages} {250605} (\bibinfo {year} {2019})}\BibitemShut {NoStop}%
\bibitem [{\citenamefont {Else}\ \emph {et~al.}(2020)\citenamefont {Else},
  \citenamefont {Machado}, \citenamefont {Nayak},\ and\ \citenamefont
  {Yao}}]{Else2020}%
  \BibitemOpen
  \bibfield  {author} {\bibinfo {author} {\bibfnamefont {D.~V.}\ \bibnamefont
  {Else}}, \bibinfo {author} {\bibfnamefont {F.}~\bibnamefont {Machado}},
  \bibinfo {author} {\bibfnamefont {C.}~\bibnamefont {Nayak}}, \ and\ \bibinfo
  {author} {\bibfnamefont {N.~Y.}\ \bibnamefont {Yao}},\ }\href {\doibase
  10.1103/PhysRevA.101.022333} {\bibfield  {journal} {\bibinfo  {journal}
  {Phys. Rev. A}\ }\textbf {\bibinfo {volume} {101}},\ \bibinfo {pages}
  {022333} (\bibinfo {year} {2020})}\BibitemShut {NoStop}%
\bibitem [{\citenamefont {Kuwahara}\ and\ \citenamefont
  {Saito}(2020{\natexlab{a}})}]{Kuwahara2020}%
  \BibitemOpen
  \bibfield  {author} {\bibinfo {author} {\bibfnamefont {T.}~\bibnamefont
  {Kuwahara}}\ and\ \bibinfo {author} {\bibfnamefont {K.}~\bibnamefont
  {Saito}},\ }\href {\doibase 10.1103/PhysRevX.10.031010} {\bibfield  {journal}
  {\bibinfo  {journal} {Phys. Rev. X}\ }\textbf {\bibinfo {volume} {10}},\
  \bibinfo {pages} {031010} (\bibinfo {year} {2020}{\natexlab{a}})}\BibitemShut
  {NoStop}%
\bibitem [{\citenamefont {Tran}\ \emph {et~al.}(2020)\citenamefont {Tran},
  \citenamefont {Chen}, \citenamefont {Ehrenberg}, \citenamefont {Guo},
  \citenamefont {Deshpande}, \citenamefont {Hong}, \citenamefont {Gong},
  \citenamefont {Gorshkov},\ and\ \citenamefont {Lucas}}]{Tran2020}%
  \BibitemOpen
  \bibfield  {author} {\bibinfo {author} {\bibfnamefont {M.~C.}\ \bibnamefont
  {Tran}}, \bibinfo {author} {\bibfnamefont {C.-F.}\ \bibnamefont {Chen}},
  \bibinfo {author} {\bibfnamefont {A.}~\bibnamefont {Ehrenberg}}, \bibinfo
  {author} {\bibfnamefont {A.~Y.}\ \bibnamefont {Guo}}, \bibinfo {author}
  {\bibfnamefont {A.}~\bibnamefont {Deshpande}}, \bibinfo {author}
  {\bibfnamefont {Y.}~\bibnamefont {Hong}}, \bibinfo {author} {\bibfnamefont
  {Z.-X.}\ \bibnamefont {Gong}}, \bibinfo {author} {\bibfnamefont {A.~V.}\
  \bibnamefont {Gorshkov}}, \ and\ \bibinfo {author} {\bibfnamefont
  {A.}~\bibnamefont {Lucas}},\ }\href {\doibase 10.1103/PhysRevX.10.031009}
  {\bibfield  {journal} {\bibinfo  {journal} {Phys. Rev. X}\ }\textbf {\bibinfo
  {volume} {10}},\ \bibinfo {pages} {031009} (\bibinfo {year}
  {2020})}\BibitemShut {NoStop}%
\bibitem [{\citenamefont {Kuwahara}\ and\ \citenamefont
  {Saito}(2021)}]{Kuwahara2021}%
  \BibitemOpen
  \bibfield  {author} {\bibinfo {author} {\bibfnamefont {T.}~\bibnamefont
  {Kuwahara}}\ and\ \bibinfo {author} {\bibfnamefont {K.}~\bibnamefont
  {Saito}},\ }\href {\doibase 10.1103/PhysRevLett.126.030604} {\bibfield
  {journal} {\bibinfo  {journal} {Phys. Rev. Lett.}\ }\textbf {\bibinfo
  {volume} {126}},\ \bibinfo {pages} {030604} (\bibinfo {year}
  {2021})}\BibitemShut {NoStop}%
\bibitem [{\citenamefont {Tran}\ \emph {et~al.}(2021)\citenamefont {Tran},
  \citenamefont {Guo}, \citenamefont {Baldwin}, \citenamefont {Ehrenberg},
  \citenamefont {Gorshkov},\ and\ \citenamefont {Lucas}}]{Tran2021}%
  \BibitemOpen
  \bibfield  {author} {\bibinfo {author} {\bibfnamefont {M.~C.}\ \bibnamefont
  {Tran}}, \bibinfo {author} {\bibfnamefont {A.~Y.}\ \bibnamefont {Guo}},
  \bibinfo {author} {\bibfnamefont {C.~L.}\ \bibnamefont {Baldwin}}, \bibinfo
  {author} {\bibfnamefont {A.}~\bibnamefont {Ehrenberg}}, \bibinfo {author}
  {\bibfnamefont {A.~V.}\ \bibnamefont {Gorshkov}}, \ and\ \bibinfo {author}
  {\bibfnamefont {A.}~\bibnamefont {Lucas}},\ }\href {\doibase
  10.1103/PhysRevLett.127.160401} {\bibfield  {journal} {\bibinfo  {journal}
  {Phys. Rev. Lett.}\ }\textbf {\bibinfo {volume} {127}},\ \bibinfo {pages}
  {160401} (\bibinfo {year} {2021})}\BibitemShut {NoStop}%
\bibitem [{\citenamefont {Wang}\ and\ \citenamefont
  {Hazzard}(2022)}]{Wang2022}%
  \BibitemOpen
  \bibfield  {author} {\bibinfo {author} {\bibfnamefont {Z.}~\bibnamefont
  {Wang}}\ and\ \bibinfo {author} {\bibfnamefont {K.~R.~A.}\ \bibnamefont
  {Hazzard}},\ }\href@noop {} {\enquote {\bibinfo {title} {Locality of gapped
  ground states in systems with power-law decaying interactions},}\ } (\bibinfo
  {year} {2022}),\ \bibinfo {note} {arXiv:2208.13057}\BibitemShut {NoStop}%
\bibitem [{\citenamefont {Defenu}\ \emph {et~al.}(2021)\citenamefont {Defenu},
  \citenamefont {Donner}, \citenamefont {Macr\`i}, \citenamefont {Pagano},
  \citenamefont {Ruffo},\ and\ \citenamefont {Trombettoni}}]{Defenu2021}%
  \BibitemOpen
  \bibfield  {author} {\bibinfo {author} {\bibfnamefont {N.}~\bibnamefont
  {Defenu}}, \bibinfo {author} {\bibfnamefont {T.}~\bibnamefont {Donner}},
  \bibinfo {author} {\bibfnamefont {T.}~\bibnamefont {Macr\`i}}, \bibinfo
  {author} {\bibfnamefont {G.}~\bibnamefont {Pagano}}, \bibinfo {author}
  {\bibfnamefont {S.}~\bibnamefont {Ruffo}}, \ and\ \bibinfo {author}
  {\bibfnamefont {A.}~\bibnamefont {Trombettoni}},\ }\href@noop {} {\enquote
  {\bibinfo {title} {Long-range interacting quantum systems},}\ } (\bibinfo
  {year} {2021}),\ \bibinfo {note} {arXiv:2109.01063}\BibitemShut {NoStop}%
\bibitem [{\citenamefont {Ritsch}\ \emph {et~al.}(2013)\citenamefont {Ritsch},
  \citenamefont {Domokos}, \citenamefont {Brennecke},\ and\ \citenamefont
  {Esslinger}}]{Esslinger2013}%
  \BibitemOpen
  \bibfield  {author} {\bibinfo {author} {\bibfnamefont {H.}~\bibnamefont
  {Ritsch}}, \bibinfo {author} {\bibfnamefont {P.}~\bibnamefont {Domokos}},
  \bibinfo {author} {\bibfnamefont {F.}~\bibnamefont {Brennecke}}, \ and\
  \bibinfo {author} {\bibfnamefont {T.}~\bibnamefont {Esslinger}},\ }\href
  {\doibase 10.1103/RevModPhys.85.553} {\bibfield  {journal} {\bibinfo
  {journal} {Rev. Mod. Phys.}\ }\textbf {\bibinfo {volume} {85}},\ \bibinfo
  {pages} {553} (\bibinfo {year} {2013})}\BibitemShut {NoStop}%
\bibitem [{\citenamefont {Gonz\'alez-Tudela}\ \emph {et~al.}(2015)\citenamefont
  {Gonz\'alez-Tudela}, \citenamefont {Hung}, \citenamefont {Chang},
  \citenamefont {Cirac},\ and\ \citenamefont {Kimble}}]{Tudela2015}%
  \BibitemOpen
  \bibfield  {author} {\bibinfo {author} {\bibfnamefont {A.}~\bibnamefont
  {Gonz\'alez-Tudela}}, \bibinfo {author} {\bibfnamefont {C.-L.}\ \bibnamefont
  {Hung}}, \bibinfo {author} {\bibfnamefont {D.~E.}\ \bibnamefont {Chang}},
  \bibinfo {author} {\bibfnamefont {J.~I.}\ \bibnamefont {Cirac}}, \ and\
  \bibinfo {author} {\bibfnamefont {H.~J.}\ \bibnamefont {Kimble}},\ }\href
  {https://doi.org/10.1038/nphoton.2015.54} {\bibfield  {journal} {\bibinfo
  {journal} {Nat. Photonics}\ }\textbf {\bibinfo {volume} {9}},\ \bibinfo
  {pages} {320} (\bibinfo {year} {2015})}\BibitemShut {NoStop}%
\bibitem [{\citenamefont {Arg\"uello-Luengo}\ \emph {et~al.}(2019)\citenamefont
  {Arg\"uello-Luengo}, \citenamefont {Gonz\'alez-Tudela}, \citenamefont {Shi},
  \citenamefont {Zoller},\ and\ \citenamefont {Cirac}}]{Cirac2019}%
  \BibitemOpen
  \bibfield  {author} {\bibinfo {author} {\bibfnamefont {J.}~\bibnamefont
  {Arg\"uello-Luengo}}, \bibinfo {author} {\bibfnamefont {A.}~\bibnamefont
  {Gonz\'alez-Tudela}}, \bibinfo {author} {\bibfnamefont {T.}~\bibnamefont
  {Shi}}, \bibinfo {author} {\bibfnamefont {P.}~\bibnamefont {Zoller}}, \ and\
  \bibinfo {author} {\bibfnamefont {J.~I.}\ \bibnamefont {Cirac}},\ }\href
  {\doibase 10.1038/s41586-019-1614-4} {\bibfield  {journal} {\bibinfo
  {journal} {Nature}\ }\textbf {\bibinfo {volume} {574}},\ \bibinfo {pages}
  {215} (\bibinfo {year} {2019})}\BibitemShut {NoStop}%
\bibitem [{\citenamefont {Browaeys}\ and\ \citenamefont
  {Lahaye}(2020)}]{Browaeys2020}%
  \BibitemOpen
  \bibfield  {author} {\bibinfo {author} {\bibfnamefont {A.}~\bibnamefont
  {Browaeys}}\ and\ \bibinfo {author} {\bibfnamefont {T.}~\bibnamefont
  {Lahaye}},\ }\href {https://doi.org/10.1038/s41567-019-0733-z} {\bibfield
  {journal} {\bibinfo  {journal} {Nat. Phys.}\ }\textbf {\bibinfo {volume}
  {16}},\ \bibinfo {pages} {132} (\bibinfo {year} {2020})}\BibitemShut
  {NoStop}%
\bibitem [{\citenamefont {Monroe}\ \emph {et~al.}(2021)\citenamefont {Monroe},
  \citenamefont {Campbell}, \citenamefont {Duan}, \citenamefont {Gong},
  \citenamefont {Gorshkov}, \citenamefont {Hess}, \citenamefont {Islam},
  \citenamefont {Kim}, \citenamefont {Linke}, \citenamefont {Pagano},
  \citenamefont {Richerme}, \citenamefont {Senko},\ and\ \citenamefont
  {Yao}}]{Monroe2021}%
  \BibitemOpen
  \bibfield  {author} {\bibinfo {author} {\bibfnamefont {C.}~\bibnamefont
  {Monroe}}, \bibinfo {author} {\bibfnamefont {W.~C.}\ \bibnamefont
  {Campbell}}, \bibinfo {author} {\bibfnamefont {L.-M.}\ \bibnamefont {Duan}},
  \bibinfo {author} {\bibfnamefont {Z.-X.}\ \bibnamefont {Gong}}, \bibinfo
  {author} {\bibfnamefont {A.~V.}\ \bibnamefont {Gorshkov}}, \bibinfo {author}
  {\bibfnamefont {P.~W.}\ \bibnamefont {Hess}}, \bibinfo {author}
  {\bibfnamefont {R.}~\bibnamefont {Islam}}, \bibinfo {author} {\bibfnamefont
  {K.}~\bibnamefont {Kim}}, \bibinfo {author} {\bibfnamefont {N.~M.}\
  \bibnamefont {Linke}}, \bibinfo {author} {\bibfnamefont {G.}~\bibnamefont
  {Pagano}}, \bibinfo {author} {\bibfnamefont {P.}~\bibnamefont {Richerme}},
  \bibinfo {author} {\bibfnamefont {C.}~\bibnamefont {Senko}}, \ and\ \bibinfo
  {author} {\bibfnamefont {N.~Y.}\ \bibnamefont {Yao}},\ }\href {\doibase
  10.1103/RevModPhys.93.025001} {\bibfield  {journal} {\bibinfo  {journal}
  {Rev. Mod. Phys.}\ }\textbf {\bibinfo {volume} {93}},\ \bibinfo {pages}
  {025001} (\bibinfo {year} {2021})}\BibitemShut {NoStop}%
\bibitem [{\citenamefont {Gong}\ and\ \citenamefont
  {Hamazaki}(2022)}]{Gong2022}%
  \BibitemOpen
  \bibfield  {author} {\bibinfo {author} {\bibfnamefont {Z.}~\bibnamefont
  {Gong}}\ and\ \bibinfo {author} {\bibfnamefont {R.}~\bibnamefont
  {Hamazaki}},\ }\href {\doibase 10.1142/S0217979222300079} {\bibfield
  {journal} {\bibinfo  {journal} {Int. J. Mod. Phys. B}\ }\textbf {\bibinfo
  {volume} {36}},\ \bibinfo {pages} {2230007} (\bibinfo {year}
  {2022})}\BibitemShut {NoStop}%
\bibitem [{\citenamefont {Schnyder}\ \emph {et~al.}(2008)\citenamefont
  {Schnyder}, \citenamefont {Ryu}, \citenamefont {Furusaki},\ and\
  \citenamefont {Ludwig}}]{Ryu2008}%
  \BibitemOpen
  \bibfield  {author} {\bibinfo {author} {\bibfnamefont {A.~P.}\ \bibnamefont
  {Schnyder}}, \bibinfo {author} {\bibfnamefont {S.}~\bibnamefont {Ryu}},
  \bibinfo {author} {\bibfnamefont {A.}~\bibnamefont {Furusaki}}, \ and\
  \bibinfo {author} {\bibfnamefont {A.~W.~W.}\ \bibnamefont {Ludwig}},\ }\href
  {\doibase 10.1103/PhysRevB.78.195125} {\bibfield  {journal} {\bibinfo
  {journal} {Phys. Rev. B}\ }\textbf {\bibinfo {volume} {78}},\ \bibinfo
  {pages} {195125} (\bibinfo {year} {2008})}\BibitemShut {NoStop}%
\bibitem [{\citenamefont {Kitaev}(2009)}]{Kitaev2009}%
  \BibitemOpen
  \bibfield  {author} {\bibinfo {author} {\bibfnamefont {A.}~\bibnamefont
  {Kitaev}},\ }\href {https://doi.org/10.1063/1.3149495} {\bibfield  {journal}
  {\bibinfo  {journal} {AIP Conf. Proc.}\ }\textbf {\bibinfo {volume} {1134}},\
  \bibinfo {pages} {22} (\bibinfo {year} {2009})}\BibitemShut {NoStop}%
\bibitem [{\citenamefont {Chiu}\ \emph {et~al.}(2016)\citenamefont {Chiu},
  \citenamefont {Teo}, \citenamefont {Schnyder},\ and\ \citenamefont
  {Ryu}}]{Ryu2016}%
  \BibitemOpen
  \bibfield  {author} {\bibinfo {author} {\bibfnamefont {C.-K.}\ \bibnamefont
  {Chiu}}, \bibinfo {author} {\bibfnamefont {J.~C.~Y.}\ \bibnamefont {Teo}},
  \bibinfo {author} {\bibfnamefont {A.~P.}\ \bibnamefont {Schnyder}}, \ and\
  \bibinfo {author} {\bibfnamefont {S.}~\bibnamefont {Ryu}},\ }\href {\doibase
  10.1103/RevModPhys.88.035005} {\bibfield  {journal} {\bibinfo  {journal}
  {Rev. Mod. Phys.}\ }\textbf {\bibinfo {volume} {88}},\ \bibinfo {pages}
  {035005} (\bibinfo {year} {2016})}\BibitemShut {NoStop}%
\bibitem [{\citenamefont {Vodola}\ \emph {et~al.}(2014)\citenamefont {Vodola},
  \citenamefont {Lepori}, \citenamefont {Ercolessi}, \citenamefont {Gorshkov},\
  and\ \citenamefont {Pupillo}}]{Vodola2014}%
  \BibitemOpen
  \bibfield  {author} {\bibinfo {author} {\bibfnamefont {D.}~\bibnamefont
  {Vodola}}, \bibinfo {author} {\bibfnamefont {L.}~\bibnamefont {Lepori}},
  \bibinfo {author} {\bibfnamefont {E.}~\bibnamefont {Ercolessi}}, \bibinfo
  {author} {\bibfnamefont {A.~V.}\ \bibnamefont {Gorshkov}}, \ and\ \bibinfo
  {author} {\bibfnamefont {G.}~\bibnamefont {Pupillo}},\ }\href {\doibase
  10.1103/PhysRevLett.113.156402} {\bibfield  {journal} {\bibinfo  {journal}
  {Phys. Rev. Lett.}\ }\textbf {\bibinfo {volume} {113}},\ \bibinfo {pages}
  {156402} (\bibinfo {year} {2014})}\BibitemShut {NoStop}%
\bibitem [{\citenamefont {Vodola}\ \emph {et~al.}(2015)\citenamefont {Vodola},
  \citenamefont {Lepori}, \citenamefont {Ercolessi},\ and\ \citenamefont
  {Pupillo}}]{Vodola2015}%
  \BibitemOpen
  \bibfield  {author} {\bibinfo {author} {\bibfnamefont {D.}~\bibnamefont
  {Vodola}}, \bibinfo {author} {\bibfnamefont {L.}~\bibnamefont {Lepori}},
  \bibinfo {author} {\bibfnamefont {E.}~\bibnamefont {Ercolessi}}, \ and\
  \bibinfo {author} {\bibfnamefont {G.}~\bibnamefont {Pupillo}},\ }\href
  {\doibase 10.1088/1367-2630/18/1/015001} {\bibfield  {journal} {\bibinfo
  {journal} {New J. Phys.}\ }\textbf {\bibinfo {volume} {18}},\ \bibinfo
  {pages} {015001} (\bibinfo {year} {2015})}\BibitemShut {NoStop}%
\bibitem [{\citenamefont {Patrick}\ \emph {et~al.}(2017)\citenamefont
  {Patrick}, \citenamefont {Neupert},\ and\ \citenamefont
  {Pachos}}]{Patrick2017}%
  \BibitemOpen
  \bibfield  {author} {\bibinfo {author} {\bibfnamefont {K.}~\bibnamefont
  {Patrick}}, \bibinfo {author} {\bibfnamefont {T.}~\bibnamefont {Neupert}}, \
  and\ \bibinfo {author} {\bibfnamefont {J.~K.}\ \bibnamefont {Pachos}},\
  }\href {\doibase 10.1103/PhysRevLett.118.267002} {\bibfield  {journal}
  {\bibinfo  {journal} {Phys. Rev. Lett.}\ }\textbf {\bibinfo {volume} {118}},\
  \bibinfo {pages} {267002} (\bibinfo {year} {2017})}\BibitemShut {NoStop}%
\bibitem [{\citenamefont {Lepori}\ \emph {et~al.}(2018)\citenamefont {Lepori},
  \citenamefont {Giuliano},\ and\ \citenamefont {Paganelli}}]{Lepori2018}%
  \BibitemOpen
  \bibfield  {author} {\bibinfo {author} {\bibfnamefont {L.}~\bibnamefont
  {Lepori}}, \bibinfo {author} {\bibfnamefont {D.}~\bibnamefont {Giuliano}}, \
  and\ \bibinfo {author} {\bibfnamefont {S.}~\bibnamefont {Paganelli}},\ }\href
  {\doibase 10.1103/PhysRevB.97.041109} {\bibfield  {journal} {\bibinfo
  {journal} {Phys. Rev. B}\ }\textbf {\bibinfo {volume} {97}},\ \bibinfo
  {pages} {041109} (\bibinfo {year} {2018})}\BibitemShut {NoStop}%
\bibitem [{\citenamefont {Viyuela}\ \emph {et~al.}(2018)\citenamefont
  {Viyuela}, \citenamefont {Fu},\ and\ \citenamefont
  {Martin-Delgado}}]{Viyuela2018}%
  \BibitemOpen
  \bibfield  {author} {\bibinfo {author} {\bibfnamefont {O.}~\bibnamefont
  {Viyuela}}, \bibinfo {author} {\bibfnamefont {L.}~\bibnamefont {Fu}}, \ and\
  \bibinfo {author} {\bibfnamefont {M.~A.}\ \bibnamefont {Martin-Delgado}},\
  }\href {\doibase 10.1103/PhysRevLett.120.017001} {\bibfield  {journal}
  {\bibinfo  {journal} {Phys. Rev. Lett.}\ }\textbf {\bibinfo {volume} {120}},\
  \bibinfo {pages} {017001} (\bibinfo {year} {2018})}\BibitemShut {NoStop}%
\bibitem [{\citenamefont {Francica}\ and\ \citenamefont
  {Dell'Anna}(2022)}]{Francica2022}%
  \BibitemOpen
  \bibfield  {author} {\bibinfo {author} {\bibfnamefont {G.}~\bibnamefont
  {Francica}}\ and\ \bibinfo {author} {\bibfnamefont {L.}~\bibnamefont
  {Dell'Anna}},\ }\href@noop {} {\enquote {\bibinfo {title} {Correlations,
  long-range entanglement and dynamics in long-range kitaev chains},}\ }
  (\bibinfo {year} {2022}),\ \bibinfo {note} {arXiv:2206.09688}\BibitemShut
  {NoStop}%
\bibitem [{\citenamefont {Bettles}\ \emph {et~al.}(2017)\citenamefont
  {Bettles}, \citenamefont {Min\'a\ifmmode~\check{r}\else \v{r}\fi{}},
  \citenamefont {Adams}, \citenamefont {Lesanovsky},\ and\ \citenamefont
  {Olmos}}]{Bettles2017}%
  \BibitemOpen
  \bibfield  {author} {\bibinfo {author} {\bibfnamefont {R.~J.}\ \bibnamefont
  {Bettles}}, \bibinfo {author} {\bibfnamefont {J.~c.~v.}\ \bibnamefont
  {Min\'a\ifmmode~\check{r}\else \v{r}\fi{}}}, \bibinfo {author} {\bibfnamefont
  {C.~S.}\ \bibnamefont {Adams}}, \bibinfo {author} {\bibfnamefont
  {I.}~\bibnamefont {Lesanovsky}}, \ and\ \bibinfo {author} {\bibfnamefont
  {B.}~\bibnamefont {Olmos}},\ }\href {\doibase 10.1103/PhysRevA.96.041603}
  {\bibfield  {journal} {\bibinfo  {journal} {Phys. Rev. A}\ }\textbf {\bibinfo
  {volume} {96}},\ \bibinfo {pages} {041603} (\bibinfo {year}
  {2017})}\BibitemShut {NoStop}%
\bibitem [{\citenamefont {Perczel}\ \emph {et~al.}(2017)\citenamefont
  {Perczel}, \citenamefont {Borregaard}, \citenamefont {Chang}, \citenamefont
  {Pichler}, \citenamefont {Yelin}, \citenamefont {Zoller},\ and\ \citenamefont
  {Lukin}}]{Perczel2017}%
  \BibitemOpen
  \bibfield  {author} {\bibinfo {author} {\bibfnamefont {J.}~\bibnamefont
  {Perczel}}, \bibinfo {author} {\bibfnamefont {J.}~\bibnamefont {Borregaard}},
  \bibinfo {author} {\bibfnamefont {D.~E.}\ \bibnamefont {Chang}}, \bibinfo
  {author} {\bibfnamefont {H.}~\bibnamefont {Pichler}}, \bibinfo {author}
  {\bibfnamefont {S.~F.}\ \bibnamefont {Yelin}}, \bibinfo {author}
  {\bibfnamefont {P.}~\bibnamefont {Zoller}}, \ and\ \bibinfo {author}
  {\bibfnamefont {M.~D.}\ \bibnamefont {Lukin}},\ }\href {\doibase
  10.1103/PhysRevA.96.063801} {\bibfield  {journal} {\bibinfo  {journal} {Phys.
  Rev. A}\ }\textbf {\bibinfo {volume} {96}},\ \bibinfo {pages} {063801}
  (\bibinfo {year} {2017})}\BibitemShut {NoStop}%
\bibitem [{\citenamefont {Kucsko}\ \emph {et~al.}(2018)\citenamefont {Kucsko},
  \citenamefont {Choi}, \citenamefont {Choi}, \citenamefont {Maurer},
  \citenamefont {Zhou}, \citenamefont {Landig}, \citenamefont {Sumiya},
  \citenamefont {Onoda}, \citenamefont {Isoya}, \citenamefont {Jelezko},
  \citenamefont {Demler}, \citenamefont {Yao},\ and\ \citenamefont
  {Lukin}}]{Kucsko2018}%
  \BibitemOpen
  \bibfield  {author} {\bibinfo {author} {\bibfnamefont {G.}~\bibnamefont
  {Kucsko}}, \bibinfo {author} {\bibfnamefont {S.}~\bibnamefont {Choi}},
  \bibinfo {author} {\bibfnamefont {J.}~\bibnamefont {Choi}}, \bibinfo {author}
  {\bibfnamefont {P.~C.}\ \bibnamefont {Maurer}}, \bibinfo {author}
  {\bibfnamefont {H.}~\bibnamefont {Zhou}}, \bibinfo {author} {\bibfnamefont
  {R.}~\bibnamefont {Landig}}, \bibinfo {author} {\bibfnamefont
  {H.}~\bibnamefont {Sumiya}}, \bibinfo {author} {\bibfnamefont
  {S.}~\bibnamefont {Onoda}}, \bibinfo {author} {\bibfnamefont
  {J.}~\bibnamefont {Isoya}}, \bibinfo {author} {\bibfnamefont
  {F.}~\bibnamefont {Jelezko}}, \bibinfo {author} {\bibfnamefont
  {E.}~\bibnamefont {Demler}}, \bibinfo {author} {\bibfnamefont {N.~Y.}\
  \bibnamefont {Yao}}, \ and\ \bibinfo {author} {\bibfnamefont {M.~D.}\
  \bibnamefont {Lukin}},\ }\href {\doibase 10.1103/PhysRevLett.121.023601}
  {\bibfield  {journal} {\bibinfo  {journal} {Phys. Rev. Lett.}\ }\textbf
  {\bibinfo {volume} {121}},\ \bibinfo {pages} {023601} (\bibinfo {year}
  {2018})}\BibitemShut {NoStop}%
\bibitem [{\citenamefont {Wahl}\ \emph {et~al.}(2013)\citenamefont {Wahl},
  \citenamefont {Tu}, \citenamefont {Schuch},\ and\ \citenamefont
  {Cirac}}]{Wahl2013}%
  \BibitemOpen
  \bibfield  {author} {\bibinfo {author} {\bibfnamefont {T.~B.}\ \bibnamefont
  {Wahl}}, \bibinfo {author} {\bibfnamefont {H.-H.}\ \bibnamefont {Tu}},
  \bibinfo {author} {\bibfnamefont {N.}~\bibnamefont {Schuch}}, \ and\ \bibinfo
  {author} {\bibfnamefont {J.~I.}\ \bibnamefont {Cirac}},\ }\href {\doibase
  10.1103/PhysRevLett.111.236805} {\bibfield  {journal} {\bibinfo  {journal}
  {Phys. Rev. Lett.}\ }\textbf {\bibinfo {volume} {111}},\ \bibinfo {pages}
  {236805} (\bibinfo {year} {2013})}\BibitemShut {NoStop}%
\bibitem [{\citenamefont {Wahl}\ \emph {et~al.}(2014)\citenamefont {Wahl},
  \citenamefont {Ha\ss{}ler}, \citenamefont {Tu}, \citenamefont {Cirac},\ and\
  \citenamefont {Schuch}}]{Wahl2014}%
  \BibitemOpen
  \bibfield  {author} {\bibinfo {author} {\bibfnamefont {T.~B.}\ \bibnamefont
  {Wahl}}, \bibinfo {author} {\bibfnamefont {S.~T.}\ \bibnamefont
  {Ha\ss{}ler}}, \bibinfo {author} {\bibfnamefont {H.-H.}\ \bibnamefont {Tu}},
  \bibinfo {author} {\bibfnamefont {J.~I.}\ \bibnamefont {Cirac}}, \ and\
  \bibinfo {author} {\bibfnamefont {N.}~\bibnamefont {Schuch}},\ }\href
  {\doibase 10.1103/PhysRevB.90.115133} {\bibfield  {journal} {\bibinfo
  {journal} {Phys. Rev. B}\ }\textbf {\bibinfo {volume} {90}},\ \bibinfo
  {pages} {115133} (\bibinfo {year} {2014})}\BibitemShut {NoStop}%
\bibitem [{\citenamefont {Dubail}\ and\ \citenamefont
  {Read}(2015)}]{Dubail2015}%
  \BibitemOpen
  \bibfield  {author} {\bibinfo {author} {\bibfnamefont {J.}~\bibnamefont
  {Dubail}}\ and\ \bibinfo {author} {\bibfnamefont {N.}~\bibnamefont {Read}},\
  }\href {\doibase 10.1103/PhysRevB.92.205307} {\bibfield  {journal} {\bibinfo
  {journal} {Phys. Rev. B}\ }\textbf {\bibinfo {volume} {92}},\ \bibinfo
  {pages} {205307} (\bibinfo {year} {2015})}\BibitemShut {NoStop}%
\bibitem [{SM()}]{SM}%
  \BibitemOpen
  \href@noop {} {}\bibinfo {note} {See Supplemental Material, which includes
  Ref.~\cite{Bhatia1997}, for details.}\BibitemShut {Stop}%
\bibitem [{O1()}]{O1}%
  \BibitemOpen
  \href@noop {} {}\bibinfo {note} {By $\mathcal{O}(1)$ constants, we mean
  independence on distance or time, which is typically assumed to be
  large.}\BibitemShut {Stop}%
\bibitem [{\citenamefont {Preiss}\ \emph {et~al.}(2015)\citenamefont {Preiss},
  \citenamefont {Ma}, \citenamefont {Tai}, \citenamefont {Lukin}, \citenamefont
  {Rispoli}, \citenamefont {Zupancic}, \citenamefont {Lahini}, \citenamefont
  {Islam},\ and\ \citenamefont {Greiner}}]{Greiner2015b}%
  \BibitemOpen
  \bibfield  {author} {\bibinfo {author} {\bibfnamefont {P.~M.}\ \bibnamefont
  {Preiss}}, \bibinfo {author} {\bibfnamefont {R.}~\bibnamefont {Ma}}, \bibinfo
  {author} {\bibfnamefont {M.~E.}\ \bibnamefont {Tai}}, \bibinfo {author}
  {\bibfnamefont {A.}~\bibnamefont {Lukin}}, \bibinfo {author} {\bibfnamefont
  {M.}~\bibnamefont {Rispoli}}, \bibinfo {author} {\bibfnamefont
  {P.}~\bibnamefont {Zupancic}}, \bibinfo {author} {\bibfnamefont
  {Y.}~\bibnamefont {Lahini}}, \bibinfo {author} {\bibfnamefont
  {R.}~\bibnamefont {Islam}}, \ and\ \bibinfo {author} {\bibfnamefont
  {M.}~\bibnamefont {Greiner}},\ }\href {\doibase 10.1126/science.1260364}
  {\bibfield  {journal} {\bibinfo  {journal} {Science}\ }\textbf {\bibinfo
  {volume} {347}},\ \bibinfo {pages} {1229} (\bibinfo {year}
  {2015})}\BibitemShut {NoStop}%
\bibitem [{PA()}]{PA}%
  \BibitemOpen
  \href@noop {} {}\bibinfo {note} {Here the ordered matrix product
  $\overleftarrow{\prod}^n_{m=1}M_m$ means $M_nM_{n-1}\cdots M_2 M_1$ (with
  $M_m=H_{\rm lr}e^{-iH_{\rm sr}(t_m - t_{m-1})}$ in Eq.~(\ref{ipte})) when
  $n>0$ and $\mathbb{1}$ when $n=0$.}\BibitemShut {Stop}%
\bibitem [{\citenamefont {Kato}(1966)}]{Kato1966}%
  \BibitemOpen
  \bibfield  {author} {\bibinfo {author} {\bibfnamefont {T.}~\bibnamefont
  {Kato}},\ }\href@noop {} {\emph {\bibinfo {title} {Perturbation Theory for
  Linear Operators}}}\ (\bibinfo  {publisher} {Springer},\ \bibinfo {address}
  {New York},\ \bibinfo {year} {1966})\BibitemShut {NoStop}%
\bibitem [{\citenamefont {Cheianov}\ and\ \citenamefont
  {Fal'ko}(2006)}]{Cheianov2006}%
  \BibitemOpen
  \bibfield  {author} {\bibinfo {author} {\bibfnamefont {V.~V.}\ \bibnamefont
  {Cheianov}}\ and\ \bibinfo {author} {\bibfnamefont {V.~I.}\ \bibnamefont
  {Fal'ko}},\ }\href {\doibase 10.1103/PhysRevLett.97.226801} {\bibfield
  {journal} {\bibinfo  {journal} {Phys. Rev. Lett.}\ }\textbf {\bibinfo
  {volume} {97}},\ \bibinfo {pages} {226801} (\bibinfo {year}
  {2006})}\BibitemShut {NoStop}%
\bibitem [{\citenamefont {Brey}\ \emph {et~al.}(2007)\citenamefont {Brey},
  \citenamefont {Fertig},\ and\ \citenamefont {Das~Sarma}}]{Brey2007}%
  \BibitemOpen
  \bibfield  {author} {\bibinfo {author} {\bibfnamefont {L.}~\bibnamefont
  {Brey}}, \bibinfo {author} {\bibfnamefont {H.~A.}\ \bibnamefont {Fertig}}, \
  and\ \bibinfo {author} {\bibfnamefont {S.}~\bibnamefont {Das~Sarma}},\ }\href
  {\doibase 10.1103/PhysRevLett.99.116802} {\bibfield  {journal} {\bibinfo
  {journal} {Phys. Rev. Lett.}\ }\textbf {\bibinfo {volume} {99}},\ \bibinfo
  {pages} {116802} (\bibinfo {year} {2007})}\BibitemShut {NoStop}%
\bibitem [{\citenamefont {Watanabe}(2018)}]{Watanabe2018}%
  \BibitemOpen
  \bibfield  {author} {\bibinfo {author} {\bibfnamefont {H.}~\bibnamefont
  {Watanabe}},\ }\href {\doibase 10.1103/PhysRevB.98.155137} {\bibfield
  {journal} {\bibinfo  {journal} {Phys. Rev. B}\ }\textbf {\bibinfo {volume}
  {98}},\ \bibinfo {pages} {155137} (\bibinfo {year} {2018})}\BibitemShut
  {NoStop}%
\bibitem [{\citenamefont {Turner}\ \emph {et~al.}(2010)\citenamefont {Turner},
  \citenamefont {Zhang},\ and\ \citenamefont {Vishwanath}}]{Vishwanath2010}%
  \BibitemOpen
  \bibfield  {author} {\bibinfo {author} {\bibfnamefont {A.~M.}\ \bibnamefont
  {Turner}}, \bibinfo {author} {\bibfnamefont {Y.}~\bibnamefont {Zhang}}, \
  and\ \bibinfo {author} {\bibfnamefont {A.}~\bibnamefont {Vishwanath}},\
  }\href {\doibase 10.1103/PhysRevB.82.241102} {\bibfield  {journal} {\bibinfo
  {journal} {Phys. Rev. B}\ }\textbf {\bibinfo {volume} {82}},\ \bibinfo
  {pages} {241102(R)} (\bibinfo {year} {2010})}\BibitemShut {NoStop}%
\bibitem [{\citenamefont {Hughes}\ \emph {et~al.}(2011)\citenamefont {Hughes},
  \citenamefont {Prodan},\ and\ \citenamefont {Bernevig}}]{Hughes2011}%
  \BibitemOpen
  \bibfield  {author} {\bibinfo {author} {\bibfnamefont {T.~L.}\ \bibnamefont
  {Hughes}}, \bibinfo {author} {\bibfnamefont {E.}~\bibnamefont {Prodan}}, \
  and\ \bibinfo {author} {\bibfnamefont {B.~A.}\ \bibnamefont {Bernevig}},\
  }\href {\doibase 10.1103/PhysRevB.83.245132} {\bibfield  {journal} {\bibinfo
  {journal} {Phys. Rev. B}\ }\textbf {\bibinfo {volume} {83}},\ \bibinfo
  {pages} {245132} (\bibinfo {year} {2011})}\BibitemShut {NoStop}%
\bibitem [{\citenamefont {Slager}\ \emph {et~al.}(2015)\citenamefont {Slager},
  \citenamefont {Rademaker}, \citenamefont {Zaanen},\ and\ \citenamefont
  {Balents}}]{Slager2015}%
  \BibitemOpen
  \bibfield  {author} {\bibinfo {author} {\bibfnamefont {R.-J.}\ \bibnamefont
  {Slager}}, \bibinfo {author} {\bibfnamefont {L.}~\bibnamefont {Rademaker}},
  \bibinfo {author} {\bibfnamefont {J.}~\bibnamefont {Zaanen}}, \ and\ \bibinfo
  {author} {\bibfnamefont {L.}~\bibnamefont {Balents}},\ }\href {\doibase
  10.1103/PhysRevB.92.085126} {\bibfield  {journal} {\bibinfo  {journal} {Phys.
  Rev. B}\ }\textbf {\bibinfo {volume} {92}},\ \bibinfo {pages} {085126}
  (\bibinfo {year} {2015})}\BibitemShut {NoStop}%
\bibitem [{\citenamefont {Hastings}(2004{\natexlab{b}})}]{Hastings2004b}%
  \BibitemOpen
  \bibfield  {author} {\bibinfo {author} {\bibfnamefont {M.~B.}\ \bibnamefont
  {Hastings}},\ }\href {\doibase 10.1103/PhysRevLett.93.126402} {\bibfield
  {journal} {\bibinfo  {journal} {Phys. Rev. Lett.}\ }\textbf {\bibinfo
  {volume} {93}},\ \bibinfo {pages} {126402} (\bibinfo {year}
  {2004}{\natexlab{b}})}\BibitemShut {NoStop}%
\bibitem [{\citenamefont {Hern\'andez-Santana}\ \emph
  {et~al.}(2017)\citenamefont {Hern\'andez-Santana}, \citenamefont {Gogolin},
  \citenamefont {Cirac},\ and\ \citenamefont {Ac\'{\i}n}}]{Hernandez2017}%
  \BibitemOpen
  \bibfield  {author} {\bibinfo {author} {\bibfnamefont {S.}~\bibnamefont
  {Hern\'andez-Santana}}, \bibinfo {author} {\bibfnamefont {C.}~\bibnamefont
  {Gogolin}}, \bibinfo {author} {\bibfnamefont {J.~I.}\ \bibnamefont {Cirac}},
  \ and\ \bibinfo {author} {\bibfnamefont {A.}~\bibnamefont {Ac\'{\i}n}},\
  }\href {\doibase 10.1103/PhysRevLett.119.110601} {\bibfield  {journal}
  {\bibinfo  {journal} {Phys. Rev. Lett.}\ }\textbf {\bibinfo {volume} {119}},\
  \bibinfo {pages} {110601} (\bibinfo {year} {2017})}\BibitemShut {NoStop}%
\bibitem [{\citenamefont {Gong}\ and\ \citenamefont {Guaita}(2021)}]{Gong2021}%
  \BibitemOpen
  \bibfield  {author} {\bibinfo {author} {\bibfnamefont {Z.}~\bibnamefont
  {Gong}}\ and\ \bibinfo {author} {\bibfnamefont {T.}~\bibnamefont {Guaita}},\
  }\href@noop {} {\enquote {\bibinfo {title} {Topology of quantum gaussian
  states and operations},}\ } (\bibinfo {year} {2021}),\ \bibinfo {note}
  {arXiv:2106.05044}\BibitemShut {NoStop}%
\bibitem [{\citenamefont {Katsura}\ and\ \citenamefont
  {Koma}(2018)}]{Katsura2018}%
  \BibitemOpen
  \bibfield  {author} {\bibinfo {author} {\bibfnamefont {H.}~\bibnamefont
  {Katsura}}\ and\ \bibinfo {author} {\bibfnamefont {T.}~\bibnamefont {Koma}},\
  }\href {\doibase 10.1063/1.5026964} {\bibfield  {journal} {\bibinfo
  {journal} {J. Math. Phys.}\ }\textbf {\bibinfo {volume} {59}},\ \bibinfo
  {pages} {031903} (\bibinfo {year} {2018})}\BibitemShut {NoStop}%
\bibitem [{\citenamefont {Lee}\ \emph {et~al.}(2019)\citenamefont {Lee},
  \citenamefont {Ahn}, \citenamefont {Zhou},\ and\ \citenamefont
  {Vishwanath}}]{Lee2019}%
  \BibitemOpen
  \bibfield  {author} {\bibinfo {author} {\bibfnamefont {J.~Y.}\ \bibnamefont
  {Lee}}, \bibinfo {author} {\bibfnamefont {J.}~\bibnamefont {Ahn}}, \bibinfo
  {author} {\bibfnamefont {H.}~\bibnamefont {Zhou}}, \ and\ \bibinfo {author}
  {\bibfnamefont {A.}~\bibnamefont {Vishwanath}},\ }\href {\doibase
  10.1103/PhysRevLett.123.206404} {\bibfield  {journal} {\bibinfo  {journal}
  {Phys. Rev. Lett.}\ }\textbf {\bibinfo {volume} {123}},\ \bibinfo {pages}
  {206404} (\bibinfo {year} {2019})}\BibitemShut {NoStop}%
\bibitem [{APB()}]{APBC}%
  \BibitemOpen
  \href@noop {} {}\bibinfo {note} {We use anti-periodic rather than periodic
  boundary conditions to avoid finite-size singularities of $h_{\rm
  fD}(\boldsymbol{k})$ at high-symmetry momenta.}\BibitemShut {Stop}%
\bibitem [{\citenamefont {Jones}\ \emph {et~al.}(2022)\citenamefont {Jones},
  \citenamefont {Thorngren},\ and\ \citenamefont {Verresen}}]{Jones2022}%
  \BibitemOpen
  \bibfield  {author} {\bibinfo {author} {\bibfnamefont {N.~G.}\ \bibnamefont
  {Jones}}, \bibinfo {author} {\bibfnamefont {R.}~\bibnamefont {Thorngren}}, \
  and\ \bibinfo {author} {\bibfnamefont {R.}~\bibnamefont {Verresen}},\
  }\href@noop {} {\enquote {\bibinfo {title} {Bulk-boundary correspondence and
  singularity-filling in long-range free-fermion chains},}\ } (\bibinfo {year}
  {2022}),\ \bibinfo {note} {arXiv:2211.15690}\BibitemShut {NoStop}%
\bibitem [{\citenamefont {Kuwahara}\ and\ \citenamefont
  {Saito}(2020{\natexlab{b}})}]{Kuwahara2020a}%
  \BibitemOpen
  \bibfield  {author} {\bibinfo {author} {\bibfnamefont {T.}~\bibnamefont
  {Kuwahara}}\ and\ \bibinfo {author} {\bibfnamefont {K.}~\bibnamefont
  {Saito}},\ }\href {https://doi.org/10.1038/s41467-020-18055-x} {\bibfield
  {journal} {\bibinfo  {journal} {Nat. Commun.}\ }\textbf {\bibinfo {volume}
  {11}},\ \bibinfo {pages} {4478} (\bibinfo {year}
  {2020}{\natexlab{b}})}\BibitemShut {NoStop}%
\bibitem [{\citenamefont {Fr\"ohlich}\ and\ \citenamefont
  {Spencer}(1983)}]{Frohlich1983}%
  \BibitemOpen
  \bibfield  {author} {\bibinfo {author} {\bibfnamefont {J.}~\bibnamefont
  {Fr\"ohlich}}\ and\ \bibinfo {author} {\bibfnamefont {T.}~\bibnamefont
  {Spencer}},\ }\href {\doibase 10.1007/BF01209475} {\bibfield  {journal}
  {\bibinfo  {journal} {Comm. Math. Phys.}\ }\textbf {\bibinfo {volume} {88}},\
  \bibinfo {pages} {151} (\bibinfo {year} {1983})}\BibitemShut {NoStop}%
\bibitem [{\citenamefont {Aizenman}\ and\ \citenamefont
  {Molchanov}(1993)}]{Aizenman1993}%
  \BibitemOpen
  \bibfield  {author} {\bibinfo {author} {\bibfnamefont {M.}~\bibnamefont
  {Aizenman}}\ and\ \bibinfo {author} {\bibfnamefont {S.}~\bibnamefont
  {Molchanov}},\ }\href {\doibase 10.1007/BF02099760} {\bibfield  {journal}
  {\bibinfo  {journal} {Comm. Math. Phys.}\ }\textbf {\bibinfo {volume}
  {157}},\ \bibinfo {pages} {245} (\bibinfo {year} {1993})}\BibitemShut
  {NoStop}%
\bibitem [{mes()}]{mesh}%
  \BibitemOpen
  \href@noop {} {}\bibinfo {note} {Given $\boldsymbol{r}=[n_1,n_2,...,n_d]$ and
  $\boldsymbol{r}'=[m_1,m_2,...,m_d]$, we can adjust the ``mesh" of
  coarse-graining along all the $d$ directions such that the difference between
  the $j$th components of $\boldsymbol{R}$ and $\boldsymbol{R}'$ is given by
  $\lceil |n_j-m_j|/\chi \rceil$. Accordingly, Eq.~(\ref{Rr}) follows with
  $R_0=\sqrt{d}$.}\BibitemShut {Stop}%
\bibitem [{\citenamefont {Bhatia}(1997)}]{Bhatia1997}%
  \BibitemOpen
  \bibfield  {author} {\bibinfo {author} {\bibfnamefont {R.}~\bibnamefont
  {Bhatia}},\ }\href@noop {} {\emph {\bibinfo {title} {Matrix Analysis}}}\
  (\bibinfo  {publisher} {Springer, New York},\ \bibinfo {year}
  {1997})\BibitemShut {NoStop}%
\end{thebibliography}%

\clearpage
\setcounter{secnumdepth}{2}
\onecolumngrid
\begin{center}
\textbf{\large Supplemental Material}
\end{center}
\setcounter{equation}{0}
\setcounter{figure}{0}
\setcounter{table}{0}
\setcounter{section}{0}
\makeatletter
\renewcommand{\theequation}{S\arabic{equation}}
\renewcommand{\thefigure}{S\arabic{figure}}
\renewcommand{\bibnumfmt}[1]{[S#1]}

We show that the generalizations to the free-fermion systems without particle-number conservation is straightforward. We then provide the full details of deriving the main results in the main text, including the Lieb-Robinson bound, clustering properties and the results related to topological phases. 

\section{Generalization to the cases without particle-number conservation}

The results presented in this paper can be generalized easily to the case of a generic free fermionic Hamiltonian with no particle number symmetry. In this case it is convenient to introduce the Majorana operators
\begin{equation}
    \hat\gamma_{\boldsymbol{r}s+}=\hat c_{\boldsymbol{r}s}^\dag + \hat c_{\boldsymbol{r}s},\;\;\;\;
    \hat\gamma_{\boldsymbol{r}s-}=i(\hat c_{\boldsymbol{r}s}^\dag - \hat c_{\boldsymbol{r}s}).
\end{equation}
For convenience, we will bunch the indices $s$ and $+/-$ of these operators into a single index $S\equiv(s,\pm)\in \tilde I=I\times\{\pm\}$. So, in short, for every lattice site $\boldsymbol{r}$ we have $2|I|$ Hermitian operators $\hat\gamma_{\boldsymbol{r}S}$, satisfying $\{\hat\gamma_{\boldsymbol{r}S},\hat\gamma_{\boldsymbol{r}'S'}\}=2\delta_{\boldsymbol{r}\boldsymbol{r}'}\delta_{SS'}$.

In this formalism, the most general free Hamiltonian is given by
\begin{equation}
\hat H = \frac{i}{2}\sum_{\substack{\boldsymbol{r},\boldsymbol{r}'\in\Lambda \\ S,S'\in \tilde I}}[A_{\boldsymbol{r}\boldsymbol{r}'}]_{SS'}\, \hat\gamma_{\boldsymbol{r}S}\hat\gamma_{\boldsymbol{r}'S'}
\label{HA}
\end{equation}
where $A_{\boldsymbol{r}\boldsymbol{r}'}$ is $2|I|\times 2|I|$ real anti-symmetric matrix. In the case of a long-range system, the $\alpha$-decay condition is represented by
\begin{equation}
\|A_{\boldsymbol{r}\boldsymbol{r}'}\|\le \frac{a}{(|\boldsymbol{r} - \boldsymbol{r}'| + 1)^\alpha}\,,
\label{eq:alphadecayA}
\end{equation}
where $a$ is an $\mathcal{O}(1)$ constant.

For a matrix satisfying~\eqref{eq:alphadecayA} with $\alpha>d$ one can apply the proof of Theorem~\ref{Thm:LR} in the main text to obtain the Lieb-Robinson bound
\begin{equation}
\|P_{\boldsymbol{r}} e^{At} P_{\boldsymbol{r}'}\| \le 
\frac{K(t)}{(|\boldsymbol{r} - \boldsymbol{r}'| + 1)^\alpha},
\label{PAt}
\end{equation}
which holds for $t$ larger than an $\mathcal{O}(1)$ constant and $K(t)\propto t^{\alpha (\alpha+1)/(\alpha -d)}$ for large $t$. While the intuitive picture of single-particle propagation no longer applies, we may relate the left-hand side of Eq.~(\ref{PAt}) to the nonequal-time anti-commutator of two Majorana modes at sites $\boldsymbol{r}$ and $\boldsymbol{r}'$.

In this context the object equivalent to the Green's function is
\begin{equation}
    Y(z)\equiv{\left(z-iA\right)}^{-1}\,.
\end{equation}
Using the same filter function as in the proof of Theorem~\ref{Thm:GF} in the main text, we have that $\int^\infty_{-\infty} dt /2\pi \, f_\sigma(t) e^{At}
$ gives $Y(z)$ up to an error controllable by $\sigma$. Analogously to Theorem~2, it follows that, if $A$ is $\alpha$-decaying with $\alpha>d$, then $Y_{\boldsymbol{r}\boldsymbol{r'}}(z)$ must satisfy the clustering property~(\ref{CPGF}) for all $z$ lying outside the spectrum of $iA$:
\begin{equation}
\|Y_{\boldsymbol{r}\boldsymbol{r}'}(z)\|\le \frac{{\rm poly}(\log(|\boldsymbol{r}-\boldsymbol{r}'|+1))}{(|\boldsymbol{r}-\boldsymbol{r}'|+1)^\alpha}.
\label{CPY}
\end{equation}

Again, this property of $Y(z)$ implies the 
clustering property 
for the covariance matrix, which in the case of no particle number symmetry should be more generally defined as
\begin{equation}
[\Gamma_{\boldsymbol{r}\boldsymbol{r}'}]_{SS'}\equiv\frac{i}{2}\langle\Psi_0| [\hat\gamma_{\boldsymbol{r}S},\hat\gamma_{\boldsymbol{r}'S'}] |\Psi_0\rangle \,.
\end{equation}
For the ground state of the Hamiltonian~\eqref{HA}, it holds that $\Gamma=i\,\mathrm{sgn}(iA)=i(\mathbb{1}-2P_<)$, where $P_<$ is the projector on the subspace spanned by the eigenstates of $iA$ with negative eigenvalues. It follows that the ground-state covariance matrix $\Gamma$ can be related to $Y(z)$ via
\begin{equation}
\Gamma=i\mathbb{1}-\oint_{\ell_<} \frac{dz}{\pi} Y(z),
\end{equation}
where $\ell_<$ is a closed loop that encircles all eigenvalues of $iA$ on the negative real semi-axis. Following a reasoning analogous to the one for $\|C_{\boldsymbol{r}\boldsymbol{r'}}\|$, we conclude that $\|\Gamma_{\boldsymbol{r}\boldsymbol{r'}}\|$ also satisfies the clustering 
property~\eqref{CPY}.

\section{Detailed derivation of the Lieb-Robinson bound}

\begin{figure}[!b]
\begin{center}
\includegraphics[width=10cm, clip]{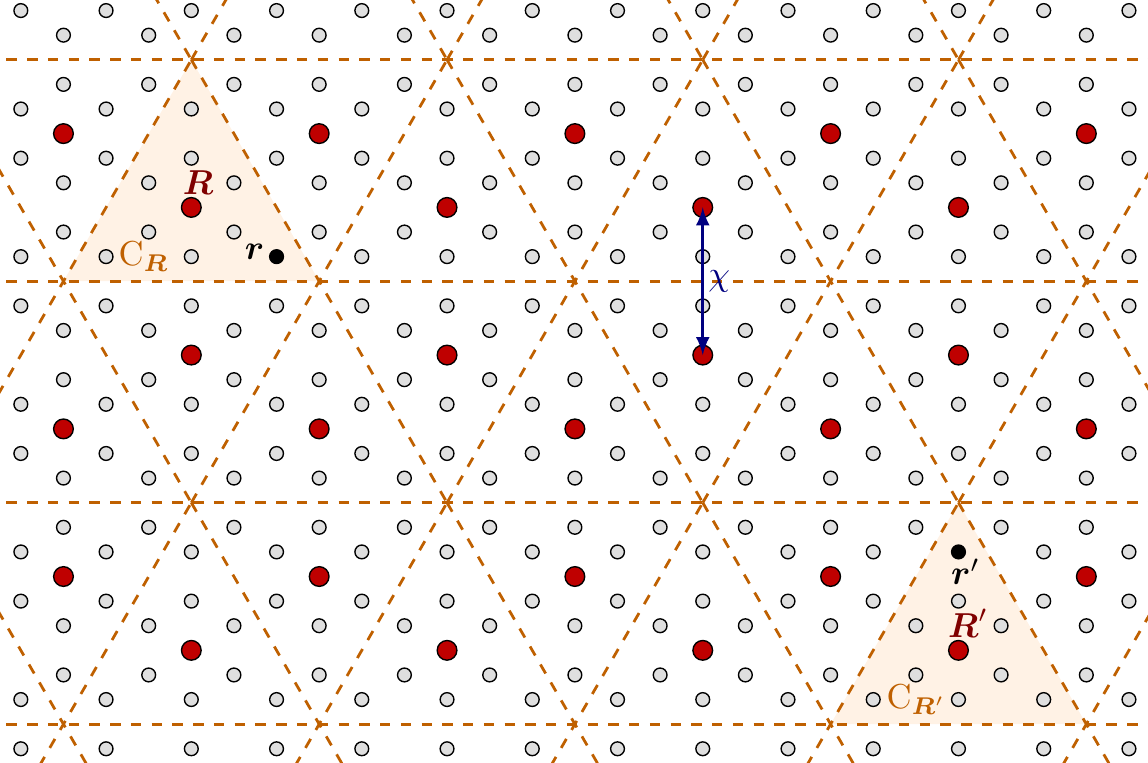}
\end{center}
   \caption{Coarse-graining of a honeycomb lattice.}
      \label{HC}
\end{figure}

Without loss of generality, we consider a $d$D lattice $\Lambda$ with unit spacing between nearest sites. We 
define its 
coarse-graining 
as another lattice $\tilde \Lambda$ which shares the same geometry as $\Lambda$ but with all the length scales multiplied by $\chi\in\mathbb{Z}^+$, which is chosen to be exactly the cutoff that separates the short-range and long-range parts in the Hamiltonian. 
Denoting ${\rm C}_{\boldsymbol{R}}\subset\Lambda$ as the set of sites coarse-grained into $\boldsymbol{R}\in\tilde\Lambda$, we have 
\begin{equation}
{\rm C}_{\boldsymbol{R}}\cap{\rm C}_{\boldsymbol{R}'}=\emptyset,\;\forall \boldsymbol{R}\neq \boldsymbol{R}',\;\;\;\;\Lambda=\cup_{\boldsymbol{R}\in\tilde\Lambda}{\rm C}_{\boldsymbol{R}},\;\;\;\; |{\rm C}_{\boldsymbol{R}}|=\chi^d.
\end{equation}
Again, the metric on $\tilde \Lambda$ is defined such that the spacing between nearest sites is unit. For any lattice geometry and coarse-graning, there always exists an $\mathcal{O}(1)$ constant $R_0$ such that $\forall \boldsymbol{r}\neq\boldsymbol{r}'\in\Lambda$
\begin{equation}
    \chi(|\boldsymbol{R}-\boldsymbol{R}'|- R_0)\le|\boldsymbol{r}-\boldsymbol{r}'|,
    \label{Rr}
\end{equation}
where $\boldsymbol{R}$ and $\boldsymbol{R}'$ are determined such that $\boldsymbol{r}\in{\rm C}_{\boldsymbol{R}}$ and $\boldsymbol{r}'\in{\rm C}_{\boldsymbol{R}'}$. For example, according to the triangle inequality of distance, we may take $R_0=2\chi^{-1}\max_{\boldsymbol{r}\in{\rm C}_{\boldsymbol{R}}}|\boldsymbol{r}-\boldsymbol{R}|_\Lambda|\sim\mathcal{O}(1)$ ($\boldsymbol{R}|_\Lambda$ means we should use the metric on $\Lambda$) as the length scale of ${\rm C}_{\boldsymbol{R}}$ is $\mathcal{O}(\chi)$. In addition, we assume that 
if we consider fixed $\boldsymbol{r}$ and $\boldsymbol{r}'$
we can properly adjust the ``mesh" of corase-graining such that 
\begin{equation}
|\boldsymbol{r}-\boldsymbol{r}'| \le \chi |\boldsymbol{R}-\boldsymbol{R}'|.
\label{Rr2}
\end{equation}
At the very least, this is achievable for the hypercubic lattice \cite{mesh}. 
See Fig.~\ref{HC} for another schematic illustration of an appropriate coarse-graining of the honeycomb lattice. 
Later we will discuss the effect of loosing Eq.~(\ref{Rr2}) to $|\boldsymbol{r}-\boldsymbol{r}'| \le \chi (|\boldsymbol{R}-\boldsymbol{R}'|+R_0)$, which always holds true (just like Eq.~(\ref{Rr})) and turns out not to alter the result qualitatively. Finally, we assume that the lattice geometry satisfies
\begin{equation}
\sum_{\boldsymbol{r}\in\Lambda:|\boldsymbol{r}|\le r}1\le b(r+1)^d,\;\;\;\;
\sum_{\boldsymbol{r}\in\Lambda:|\boldsymbol{r}|\ge r} \frac{1}{(|\boldsymbol{r}| + 1)^\alpha} \le \frac{c_1}{(r+1)^{\alpha - d}},\;\;\forall r\ge0,
\label{b1}
\end{equation}
where 
$b$ and $c_1$ are two $\mathcal{O}(1)$ constants independent of $r$.

We move on to introduce two lemmas which are crucial for bounding the Taylor-expansion terms in the interaction picture 
under coarse-graining: 
\begin{lemma}[Coarse-grained long-range decay]
\label{CGLR}
Defining a complete set of ``coarse-grained" orthogonal projectors 
$P_{\boldsymbol{R}}\equiv \sum_{\boldsymbol{r}\in {\rm C}_{\boldsymbol{R}}} P_{\boldsymbol{r}}$, 
we have
\begin{equation}
\| P_{\boldsymbol{R}} H_{\rm lr} P_{\boldsymbol{R}'}\| \le \frac{C_1\chi^{-(\alpha - d)}}{(|\boldsymbol{R} - \boldsymbol{R}'| + 1)^\alpha}.
\label{Hld}
\end{equation}
Here $C_1=c_1 J (R_0 + 1 + c_1^{-1/\alpha})^\alpha$ is an $\mathcal{O}(1)$ constant independent of 
$\chi$.
\end{lemma}
\emph{Proof}: We first note that there is always a trivial upper bound given by the operator norm of $H_{\rm lr}$:
\begin{equation}
\| P_{\boldsymbol{R}} H_{\rm lr} P_{\boldsymbol{R}'}\| \le \|H_{\rm lr}\| \le \max_{\boldsymbol{r}\in\Lambda}\sum_{\boldsymbol{r}'\in\Lambda} \|[H_{\rm lr}]_{\textcolor{black}{\boldsymbol{r}\boldsymbol{r}'}}\|
\le  \max_{\boldsymbol{r}\in\Lambda}\sum_{\boldsymbol{r}'\in\Lambda: |\boldsymbol{r}'-\boldsymbol{r}|>\chi}\frac{J}{(|\boldsymbol{r}-\boldsymbol{r}'|+1)^\alpha} \le \frac{c_1J}{(\chi +1)^{\alpha-d}}
< c_1 J\chi^{-(\alpha - d)}.
\label{Hlt}
\end{equation}
Here the second inequality arises from:
\begin{equation}
\begin{split}
\|H_{\rm lr}\| &=|\boldsymbol{\psi}^\dag H_{\rm lr} \boldsymbol{\psi}|\le \sum_{\boldsymbol{r},\boldsymbol{r}'\in\Lambda} |\boldsymbol{\psi}^\dag P_{\boldsymbol{r}} H_{\rm lr} P_{\boldsymbol{r}'}\boldsymbol{\psi}|
\le \sum_{\boldsymbol{r},\boldsymbol{r}'\in\Lambda} \|\boldsymbol{\psi}^\dag P_{\boldsymbol{r}}\| \| P_{\boldsymbol{r}}H_{\rm lr} P_{\boldsymbol{r}'}\| \| P_{\boldsymbol{r}'}\boldsymbol{\psi}\|  \\
&\le  \sum_{\boldsymbol{r},\boldsymbol{r}'\in\Lambda} \frac{1}{2}(\boldsymbol{\psi}^\dag P_{\boldsymbol{r}}\boldsymbol{\psi} + \boldsymbol{\psi}^\dag P_{\boldsymbol{r}'}\boldsymbol{\psi}) \| P_{\boldsymbol{r}}H_{\rm lr} P_{\boldsymbol{r}'}\| 
= \sum_{\boldsymbol{r},\boldsymbol{r}'\in\Lambda}\boldsymbol{\psi}^\dag P_{\boldsymbol{r}}\boldsymbol{\psi} \| P_{\boldsymbol{r}}H_{\rm lr} P_{\boldsymbol{r}'}\| \le \max_{\boldsymbol{r}\in\Lambda} \sum_{\boldsymbol{r}'\in\Lambda:|\boldsymbol{r}'-\boldsymbol{r}|>\chi}\| H_{\textcolor{black}{\boldsymbol{r}\boldsymbol{r}'}}\|,
\label{Hnorm}
\end{split}
\end{equation}
where $\boldsymbol{\psi}$ is a normalized eigenvector of $H_{\rm lr}$ with the largest (absolute) eigenvalue and the identity $\sum_{\boldsymbol{r}\in\Lambda}\boldsymbol{\psi}^\dag P_{\boldsymbol{r}}\boldsymbol{\psi}=1$  and $\boldsymbol{\psi}^\dag P_{\boldsymbol{r}}\boldsymbol{\psi}\ge0$ have been used. To make use of the distance between $\boldsymbol{R}$ and $\boldsymbol{R}'$, we follow a similar procedure as Eq.~(\ref{Hnorm}):
\begin{equation}
\begin{split}
\| P_{\boldsymbol{R}} H_{\rm lr} P_{\boldsymbol{R}'}\|  &= |\boldsymbol{\varphi}^\dag P_{\boldsymbol{R}}H_{\rm lr} P_{\boldsymbol{R}'}  \boldsymbol{\varphi}'| 
\le \sum_{\boldsymbol{r}\in {\rm C}_{\boldsymbol{R}},\boldsymbol{r}'\in{\rm C}_{\boldsymbol{R}'}}
\frac{1}{2}(\boldsymbol{\varphi}^\dag P_{\boldsymbol{r}} \boldsymbol{\varphi } + \boldsymbol{\varphi}'^\dag P_{\boldsymbol{r}'} \boldsymbol{\varphi}') \|[H_{\rm lr}]_{\textcolor{black}{\boldsymbol{r}\boldsymbol{r}'}}\| \\
&\le \frac{J}{[(|\boldsymbol{R}-\boldsymbol{R}'|-R_0)\chi+1]^\alpha} \sum_{\boldsymbol{r}\in {\rm C}_{\boldsymbol{R}},\boldsymbol{r}'\in{\rm C}_{\boldsymbol{R}'}}
\frac{1}{2}(\boldsymbol{\varphi}^\dag P_{\boldsymbol{r}} \boldsymbol{\varphi} + \boldsymbol{\varphi}'^\dag P_{\boldsymbol{r}'} \boldsymbol{\varphi}')
=\frac{J|C_{\boldsymbol{R}}|}{[(|\boldsymbol{R}-\boldsymbol{R}'|-R_0)\chi+1]^\alpha} \\
&\le \frac{J\chi^d}{[(|\boldsymbol{R}-\boldsymbol{R}'|-R_0)\chi+1]^\alpha} < \frac{J\chi^{-(\alpha-d)}}{(|\boldsymbol{R}-\boldsymbol{R}'|-R_0)^\alpha},
\end{split}
\label{Hln}
\end{equation}
where $\boldsymbol{\varphi}=P_{\boldsymbol{R}}\boldsymbol{\varphi}$ and $\boldsymbol{\varphi}'=P_{\boldsymbol{R}'}\boldsymbol{\varphi}'$ are two normalized vectors that validate the first equality and Eq.~(\ref{Rr}) has been used. Now suppose $|\boldsymbol{R}-\boldsymbol{R}'|< R_0 + c_1^{-1/\alpha}$, then Eq.~(\ref{Hld}) follows from the trivial bound (\ref{Hlt}). Otherwise, $|\boldsymbol{R}-\boldsymbol{R}'|\ge R_0 + c_1^{-1/\alpha}$, one can check that
\begin{equation}
\frac{1}{(|\boldsymbol{R}-\boldsymbol{R}'|-R_0)^\alpha} \le \frac{c_1(R_0 + 1 +c_1^{-\frac{1}{\alpha}})^\alpha}{(|\boldsymbol{R}-\boldsymbol{R}'|+1)^\alpha}, 
\end{equation}
so Eq.~(\ref{Hld}) follows from Eq.~(\ref{Hln}). \hfill$\Box$

\begin{lemma}[Coarse-grained short-range Lieb-Robinson bound]
\label{CGSRLR}
Following the same notations in Lemma~\ref{CGLR}, we have
\begin{equation}
\| P_{\boldsymbol{R}} e^{-iH_{\rm sr} t} P_{\boldsymbol{R}'}\| \le \min\{C_2 e^{vt - |\boldsymbol{R} - \boldsymbol{R}'|},1\}.
\label{RLR}
\end{equation}
Here $v= e c_1 J$ and $C_2=e^{R_0}$ are both $\mathcal{O}(1)$ constants independent of $\chi$.
\end{lemma}
\emph{Proof}: 
The trivial bound simply follows from the unitarity:
\begin{equation}
\| P_{\boldsymbol{R}} e^{-iH_{\rm sr} t} P_{\boldsymbol{R}'}\|\le \|e^{-iH_{\rm sr} t} \|=1.
\label{RLR1}
\end{equation}
To incorporate the distance dependence, we follow the standard estimation \cite{FossFeig2015,Gong2022}
\begin{equation}
\| P_{\boldsymbol{R}} e^{-iH_{\rm sr} t} P_{\boldsymbol{R}'}\| \le \sum^\infty_{n=\left\lceil \frac{{\rm dist}({\rm C}_{\boldsymbol{R}},{\rm C}_{\boldsymbol{R}'})}{\chi}\right\rceil} \frac{1}{n!} \| H_{\rm sr}\|^n t^n 
\le e^{e\|H_{\rm sr}\| t - \left\lceil {\rm dist}({\rm C}_{\boldsymbol{R}},{\rm C}_{\boldsymbol{R}'})/\chi\right\rceil}
\le e^{e\|H_{\rm sr}\| t - {\rm dist}({\rm C}_{\boldsymbol{R}},{\rm C}_{\boldsymbol{R}'})/\chi},
\end{equation}
where we have used the fact that the hopping range of $H_{\rm sr}$ is $\chi$ (this is why the sum starts from $\lceil {\rm dist}({\rm C}_{\boldsymbol{R}},{\rm C}_{\boldsymbol{R}'})/\chi \rceil$) as well as the inequality 
\begin{equation}
\sum^\infty_{n=m}\frac{e^mx^n}{n!}\le\sum^\infty_{n=0}\frac{(ex)^n}{n!} =e^{ex},\;\;\forall x\ge0,\;m\in\mathbb{N}. 
\end{equation}
Following a similar derivation as Eq.~(\ref{Hnorm}), we know that 
\begin{equation}
\|H_{\rm sr}\|\le \max_{\boldsymbol{r}\in\Lambda} \sum_{\boldsymbol{r}'\in\Lambda:|\boldsymbol{r}'-\boldsymbol{r}|\le\chi}\| H_{\textcolor{black}{\boldsymbol{r}\boldsymbol{r}'}}\|< \sum_{\boldsymbol{r}'\in\Lambda}\frac{J}{(|\boldsymbol{r}'-\boldsymbol{r}|+1)^\alpha}\le c_1J, 
\end{equation}
where we have used Eq.~(\ref{b1}). Further combining with Eq.~(\ref{Rr}), we obtain
\begin{equation}
\| P_{\boldsymbol{R}} e^{-iH_{\rm sr} t} P_{\boldsymbol{R}'}\| \le \max_{\boldsymbol{r}\in {\rm C}_{\boldsymbol{R}},\boldsymbol{r}'\in{\rm C}_{\boldsymbol{R}'}} e^{vt - |\boldsymbol{r} - \boldsymbol{r}'|/\chi } \le e^{R_0} e^{vt - |\boldsymbol{R} - \boldsymbol{R}'|}.
\label{RLR2}
\end{equation}
Combining Eqs.~(\ref{RLR1}) and (\ref{RLR2}), we obtain the desired result (\ref{RLR}). \hfill$\Box$

\textcolor{black}{Having the above two Lemmas in mind, we can already understand the argument for the $\chi^d$-factor improvement in the main text on a rather quantitative level. Recalling that $|\boldsymbol{r}-\boldsymbol{r}'|$ can be roughly estimated by $\chi|\boldsymbol{R}-\boldsymbol{R}'|$, we know that $\|P_{\boldsymbol{r}}H_{\rm lr}P_{\boldsymbol{r}'}\|\sim J(\chi|\boldsymbol{R}-\boldsymbol{R}'|)^{-\alpha}$ and thus $\sum_{\boldsymbol{r}\in{\rm C}_{\boldsymbol{R}},\boldsymbol{r}'\in{\rm C}_{\boldsymbol{R}'}}\|P_{\boldsymbol{r}}H_{\rm lr}P_{\boldsymbol{r}'}\|\sim J\chi^{-(\alpha-2d)}/|\boldsymbol{R}-\boldsymbol{R}'|^{\alpha}$ (since $|{\rm C}_{\boldsymbol{R}}|=|{\rm C}_{\boldsymbol{R}'}|=\chi^d$), which is indeed larger than the rhs in Eq.~(\ref{Hld}) by roughly a factor $\chi^d$. Regarding the statement that $\|P_{\boldsymbol{R}}e^{-iH_{\rm sr}t}P_{\boldsymbol{R}'}\|$ and $\|P_{\boldsymbol{r}}e^{-iH_{\rm sr}t}P_{\boldsymbol{r}'}\|$ (more precisely, their upper bounds) differ by mostly an $\mathcal{O}(1)$ factor, this is already clear in the derivation of Lemma~(\ref{CGSRLR}), especially in Eq.~(\ref{RLR2}).}

\textcolor{black}{To proceed rigorously}, we will also need the following two propositions (see also the Supplemental Material of Ref.~\cite{FossFeig2015}). For the sake of consistency, we present them in terms of the coarse-grained lattice $\tilde\Lambda$  as they will be used in this context.
\begin{proposition}[Bound on the convolution of exponential and algebraic decays] 
For any $\xi>\alpha -1$, we have
\begin{equation}
\sum_{\boldsymbol{R}\in \tilde \Lambda} \frac{\min\{e^{\xi - |\boldsymbol{R} - \boldsymbol{R}_2|},1\}}{(|\boldsymbol{R}_1 - \boldsymbol{R} |+1)^\alpha} 
\le c_1\min\left\{\left[\frac{4(\xi + 1)}{|\boldsymbol{R}_1 - \boldsymbol{R}_2 |+1}\right]^\alpha,1\right\}.
\label{eps}
\end{equation}
\label{Prop:ea}
\end{proposition}
\emph{Proof}: 
We first note that there is always a trivial bound
\begin{equation}
\sum_{\boldsymbol{R}\in \tilde \Lambda} \frac{\min\{e^{\xi - |\boldsymbol{R} - \boldsymbol{R}_2|},1\}}{(|\boldsymbol{R}_1 - \boldsymbol{R} |+1)^\alpha} \le 
\sum_{\boldsymbol{R}\in \tilde \Lambda} \frac{1}{(|\boldsymbol{R}_1 - \boldsymbol{R} |+1)^\alpha} \le c_1.
\end{equation}
The other bound is relevant in the cases in which 
$|\boldsymbol{R}_1 - \boldsymbol{R}_2| \ge 4\xi+3>4\xi$. To prove it we basically follow the idea of Ref.~\cite{FossFeig2015}, which is to split the sum into two parts:
\begin{equation}
\sum_{\boldsymbol{R}\in \tilde \Lambda} =  \sum_{\boldsymbol{R}\in \tilde \Lambda: |\boldsymbol{R} - \boldsymbol{R}_1|< |\boldsymbol{R}_1 - \boldsymbol{R}_2|/2} +  \sum_{\boldsymbol{R}\in \tilde \Lambda: |\boldsymbol{R} - \boldsymbol{R}_1|\ge |\boldsymbol{R}_1 - \boldsymbol{R}_2|/2} 
\end{equation}
For the first part, $|\boldsymbol{R} - \boldsymbol{R}_1|< |\boldsymbol{R}_1 - \boldsymbol{R}_2|/2$ necessarily implies 
\begin{equation}
|\boldsymbol{R} - \boldsymbol{R}_2| \ge |\boldsymbol{R}_1 - \boldsymbol{R}_2| - |\boldsymbol{R} - \boldsymbol{R}_1| > \frac{1}{2} |\boldsymbol{R}_1 - \boldsymbol{R}_2|.
\end{equation}
Recalling that we are considering the case in which $\xi<|\boldsymbol{R}_1 - \boldsymbol{R}_2|/4$, we have
\begin{equation}
\begin{split}
\sum_{\boldsymbol{R}\in \tilde \Lambda: |\boldsymbol{R} - \boldsymbol{R}_1|< |\boldsymbol{R}_1 - \boldsymbol{R}_2|/2} \frac{\min\{e^{\xi - |\boldsymbol{R} - \boldsymbol{R}_2|},1\}}{(|\boldsymbol{R}_1 - \boldsymbol{R} |+1)^\alpha}
&< e^{-\frac{1}{4}|\boldsymbol{R}_1 - \boldsymbol{R}_2|}\sum_{\boldsymbol{R}\in \tilde \Lambda}\frac{1}{(|\boldsymbol{R}_1 - \boldsymbol{R} |+1)^\alpha} \\
&\le c_1 e^{-\frac{1}{4}|\boldsymbol{R}_1 - \boldsymbol{R}_2|}
<\frac{c_1}{2}\left[\frac{4(\xi + 1)}{|\boldsymbol{R}_1 - \boldsymbol{R}_2 |+1}\right]^\alpha,
\end{split}
\label{ea1}
\end{equation}
where we have used the fact
\begin{equation}
e^{-\frac{1}{4}|\boldsymbol{R}_1 - \boldsymbol{R}_2|}\le \frac{e^{\frac{1}{4}-\alpha}(4\alpha)^\alpha}{(|\boldsymbol{R}_1 - \boldsymbol{R}_2|+1)^\alpha},\;\;\;\;
e^{\frac{1}{4}-\alpha}<e^{-\frac{3}{4}}=0.472...<\frac{1}{2}. 
\end{equation}
For the second part, we can bound the sum by
\begin{equation}
\begin{split}
\sum_{\boldsymbol{R}\in \tilde \Lambda: |\boldsymbol{R} - \boldsymbol{R}_1|\ge |\boldsymbol{R}_1 - \boldsymbol{R}_2|/2} \frac{\min\{e^{\xi - |\boldsymbol{R} - \boldsymbol{R}_2|},1\}}{(|\boldsymbol{R}_1 - \boldsymbol{R} |+1)^\alpha}
&< \frac{2^\alpha}{(|\boldsymbol{R}_1 - \boldsymbol{R}_2 |+1)^\alpha} \sum_{\boldsymbol{R}\in \tilde \Lambda} \min\{e^{\xi - |\boldsymbol{R} - \boldsymbol{R}_2|},1\} \\ 
&\le \left[\frac{2(\xi + 1)}{|\boldsymbol{R}_1 - \boldsymbol{R}_2 |+1}\right]^\alpha\sum_{\boldsymbol{R}\in \tilde \Lambda} \frac{1}{(|\boldsymbol{R} - \boldsymbol{R}_2|+1)^\alpha} <
\frac{c_1}{2}\left[\frac{4(\xi + 1)}{|\boldsymbol{R}_1 - \boldsymbol{R}_2 |+1}\right]^\alpha,
\end{split}
\label{ea2}
\end{equation}
where we have used the fact that, for $|\boldsymbol{R} - \boldsymbol{R}_2|\ge\xi >\alpha -1$ (since $(x+1)^\alpha e^{-x}$ decreases monotonically on $[\alpha-1,\infty)$)
\begin{equation}
(|\boldsymbol{R} - \boldsymbol{R}_2| + 1)^\alpha e^{-|\boldsymbol{R} - \boldsymbol{R}_2|}\le (\xi +1)^\alpha e^{-\xi}
\end{equation}
which implies
\begin{equation}
\min\{e^{\xi-|\boldsymbol{R} - \boldsymbol{R}_2|},1\}\le\left(\frac{\xi+1}{|\boldsymbol{R} - \boldsymbol{R}_2|+1}\right)^\alpha.
\end{equation}
Combining Eqs.~(\ref{ea1}) and (\ref{ea2}), we obtain the desired result (\ref{eps}). \hfill$\Box$
\begin{proposition}[Reproducibility] Defining $F(\boldsymbol{R})=\min\{[(\Theta+1)/(|\boldsymbol{R}|+1)]^\alpha,1\}$ with $\Theta\ge0$, we have
\begin{equation}
    \sum_{\boldsymbol{R}\in\tilde\Lambda}F(\boldsymbol{R}_1-\boldsymbol{R})F(\boldsymbol{R}-\boldsymbol{R}_2)< c_2(\Theta+1)^d F(\boldsymbol{R}_1-\boldsymbol{R}_2),
\end{equation}
where $c_2=2^{\alpha+1}(b+c_1)$ is an $\mathcal{O}(1)$ constant.
\label{Prop:rp}
\end{proposition}
\emph{Proof}: Noting that $F(\boldsymbol{R})\le1$ by definition, we have the following trivial bound (independent of $\boldsymbol{R}_{1,2}$):
\begin{equation}
\begin{split}
    \sum_{\boldsymbol{R}\in\tilde\Lambda}F(\boldsymbol{R}_1-\boldsymbol{R})F(\boldsymbol{R}-\boldsymbol{R}_2)
    &<\sum_{\boldsymbol{R}\in\tilde\Lambda}F(\boldsymbol{R}_1-\boldsymbol{R})=\sum_{\boldsymbol{R}\in\tilde\Lambda:|\boldsymbol{R}-\boldsymbol{R}_1|<R_0}1+ \sum_{\boldsymbol{R}\in\tilde\Lambda:|\boldsymbol{R}-\boldsymbol{R}_1|\ge R_0}\left(\frac{\Theta+1}{|\boldsymbol{R}-\boldsymbol{R}_1|+1}\right)^\alpha \\
    &< b (\Theta+1)^d + (R_0 +1)^\alpha\frac{c_1}{(\Theta+1)^{\alpha-d}}=(b+c_1)(\Theta+1)^d,
    \end{split}
\end{equation}
where we have used Eq.~(\ref{b1}). Using this result, we further have
\begin{equation}
\begin{split}
\sum_{\boldsymbol{R}\in\tilde\Lambda}F(\boldsymbol{R}_1-\boldsymbol{R})F(\boldsymbol{R}-\boldsymbol{R}_2)&< 
\sum_{\boldsymbol{R}\in\tilde\Lambda} F\left(\frac{\boldsymbol{R}_1-\boldsymbol{R}_2}{2}\right)[F(\boldsymbol{R}_1-\boldsymbol{R})+F(\boldsymbol{R}-\boldsymbol{R}_2)] \\
&< 2^{\alpha+1}(b+c_1)(\Theta+1)^d F(\boldsymbol{R}_1-\boldsymbol{R}_2),
    \end{split}
\end{equation}
where we have also used $\min\{F(\boldsymbol{R}_1-\boldsymbol{R}),F(\boldsymbol{R}-\boldsymbol{R}_2)\}\le F((\boldsymbol{R}_1-\boldsymbol{R}_2)/2)<2^\alpha F(\boldsymbol{R}_1-\boldsymbol{R}_2)$ and $\max\{F(\boldsymbol{R}_1-\boldsymbol{R}),F(\boldsymbol{R}-\boldsymbol{R}_2)\}<F(\boldsymbol{R}_1-\boldsymbol{R})+F(\boldsymbol{R}-\boldsymbol{R}_2)$. \hfill$\Box$ 

Having the above preliminaries in mind, we are in a position to derive the main result. 
We first apply the norm triangle inequality to each term in the Taylor expansion of $P_{\boldsymbol{r}}e^{-iHt}P_{\boldsymbol{r}'}$ in the interaction picture (cf. Eq.~(\ref{ipte})):
\begin{equation}
\| P_{\boldsymbol{r}} e^{-iHt} P_{\boldsymbol{r}'}\| \le \sum^\infty_{n=0} \int^t_{0} dt_n\int^{t_n}_0 dt_{n-1} \cdots \int^{t_2}_0 dt_1 \| P_{\boldsymbol{r}} e^{-iH_{\rm sr}(t-t_n)}H_{\rm lr}e^{-iH_{\rm sr}(t_n-t_{n-1})} \cdots H_{\rm lr}e^{-iH_{\rm sr}t_1} P_{\boldsymbol{r}'}\|.
\end{equation}
By inserting the \emph{coarse-grained} projectors, 
we can upper bound each term by (cf. Eq.~(\ref{PRB}))
\begin{equation}
\begin{split}
&\| P_{\boldsymbol{r}} e^{-iH_{\rm sr}(t-t_n)}H_{\rm lr}e^{-iH_{\rm sr}(t_n-t_{n-1})} \cdots H_{\rm lr}e^{-iH_{\rm sr}t_1} P_{\boldsymbol{r}'}\| \\
\le& \sum_{\{\boldsymbol{R}_j\in\tilde\Lambda\}^{2n}_{j=1}} \|P_{\boldsymbol{R}} e^{-iH_{\rm sr}(t-t_n)} P_{\boldsymbol{R}_{2n}}\| \prod^n_{m=1}\|P_{\boldsymbol{R}_{2m}}H_{\rm lr}P_{\boldsymbol{R}_{2m-1}} \| \|P_{\boldsymbol{R}_{2m-1}} e^{-iH_{\rm sr}(t_m-t_{m-1})}P_{\boldsymbol{R}_{2m-2}}\| \\ 
\le&C^n_1\chi^{-n(\alpha-d)} \sum_{\{\boldsymbol{R}_j\in\tilde\Lambda\}^{2n}_{j=1}} \min\{C_2e^{v(t-t_n)-|\boldsymbol{R} - \boldsymbol{R}_{2n}|},1\} \prod^n_{m=1}\frac{\min\{C_2e^{v(t_m - t_{m-1})-|\boldsymbol{R}_{2m-1} - \boldsymbol{R}_{2m-2}|},1\} }{(|\boldsymbol{R}_{2m} - \boldsymbol{R}_{2m-1}|+1)^\alpha}\\
\le&C^n_1\chi^{-n(\alpha-d)} \sum_{\{\boldsymbol{R}_j\in\tilde\Lambda\}^{2n}_{j=1}} \min\{e^{vt+R_0-|\boldsymbol{R} - \boldsymbol{R}_{2n}|},1\} \prod^n_{m=1}\frac{\min\{e^{vt+R_0-|\boldsymbol{R}_{2m-1} - \boldsymbol{R}_{2m-2}|},1\} }{(|\boldsymbol{R}_{2m} - \boldsymbol{R}_{2m-1}|+1)^\alpha},
\end{split}
\label{Ptn}
\end{equation}
where Lemmas~\ref{CGLR} and \ref{CGSRLR} have been used. To proceed, we use Propositions~\ref{Prop:ea} and  
\ref{Prop:rp} as well as 
\begin{equation}
\min \{ e^{\xi-|\boldsymbol{R}|},1 \}
\le\min\left\{\left(\frac{\xi+1}{|\boldsymbol{R}|+1}\right)^\alpha,1\right\}
\le \min\left\{\left[\frac{4(\xi+1)}{|\boldsymbol{R}|+1}\right]^\alpha,1\right\}, 
\label{eab}
\end{equation}
obtaining the following bound on Eq.~(\ref{Ptn}) for any $n\in\mathbb{N}$:
\begin{equation}
\| P_{\boldsymbol{r}} e^{-iH_{\rm sr}(t-t_n)}H_{\rm lr}e^{-iH_{\rm sr}(t_n-t_{n-1})} \cdots H_{\rm lr}e^{-iH_{\rm sr}t_1} P_{\boldsymbol{r}'}\|
\le \frac{4^\alpha(vt+R_0+1)^\alpha[c_1c_2C_1 4^d(vt+R_0+1)^d\chi^{-(\alpha - d)}]^n}{(|\boldsymbol{R} - \boldsymbol{R}'|+1)^\alpha}, 
\end{equation}
which is valid as long as $vt + R_0 > \alpha -1$, that is $t>\textcolor{black}{t_{\rm c}}
=(\alpha -1-R_0)/v$. 
Defining
\begin{equation}
\lambda_\chi = c_1c_2C_14^d(vt+R_0+1)^d\chi^{-(\alpha - d)}
\end{equation}
and summing up all the Taylor-expansion terms, we end up with
\begin{equation}
\| P_{\boldsymbol{r}} e^{-iHt} P_{\boldsymbol{r}'}\| \le \frac{4^\alpha(vt+R_0+1)^\alpha e^{\lambda_\chi t}}{(|\boldsymbol{R} - \boldsymbol{R}'|+1)^\alpha}
\le  \frac{4^\alpha\chi^\alpha(vt+R_0+1)^\alpha  e^{\lambda_\chi t} }{(|\boldsymbol{r} - \boldsymbol{r}'|+1)^\alpha},
\end{equation}
where we have used Eq.~(\ref{Rr2}). Note that $\lambda_\chi$ can be made sufficiently small by appropriately choosing a large $\chi$. In particular, requiring $\lambda_\chi t \le1$, we may choose $\chi=\lceil \chi_t\rceil$ with 
\begin{equation}
\chi_t = [c_1c_2 C_1 4^d(vt+R_0 +1)^d t]^{\frac{1}{\alpha - d}}, 
\end{equation}
which scales as $t^{(d+1)/(\alpha-d)}$ for large $t$, leading to the desired result (i.e., Eq.~(\ref{Thm:LR})):
\begin{equation}
\| P_{\boldsymbol{r}} e^{-iHt} P_{\boldsymbol{r}'}\| \le  \frac{K(t)}{(|\boldsymbol{r} - \boldsymbol{r}'|+1)^\alpha},
\label{GGCb}
\end{equation}
Here the time-dependent coefficient of the algebraic tail reads
\begin{equation}
K(t) = e[4(\chi_t+1)(vt+R_0+1)]^\alpha,
\label{Kt}
\end{equation}
which scales as $t^{\alpha(\alpha +1)/(\alpha - d)}$ for large $t$. 

Before ending the section, we would like to mention that a better analogy of the main result in Ref.~\cite{FossFeig2015} reads
\begin{equation}
\| P_{\boldsymbol{r}} e^{-iHt} P_{\boldsymbol{r}'}\| 
\le e^{vt - \frac{|\boldsymbol{r}-\boldsymbol{r}'|}{\chi_t+1}} +
\frac{(1-e^{-1})K(t)}{(|\boldsymbol{r} - \boldsymbol{r}'|+1)^\alpha}.
\label{epa}
\end{equation}
This result can be obtained by directly applying Lemma~\ref{CGSRLR} to the $0$th ($n=0$) term in the Taylor expansion instead of upper bounding it by Eq.~(\ref{eab}). However, such a small improvement has no obvious advantage in tightening other bounds but rather makes the derivations more complicated. Therefore, we prefer Eq.~(\ref{GGCb}) and will use it rather than Eq.~(\ref{epa}) hereafter. We would also like to mention that loosing Eq.~(\ref{Rr2}) to $|\boldsymbol{r}-\boldsymbol{r}'|\le\chi(|\boldsymbol{R}-\boldsymbol{R}'|+R_0)$ does not alter the result so much. For $R_0>1$, as long as we further require $\chi\le(|\boldsymbol{r}-\boldsymbol{r}'|-1)/(R_0-1)$, one can check that Eq.~(\ref{GGCb}) still holds true with $K(t)$ in Eq.~(\ref{Kt}) multiplied by a constant $2^\alpha$. The additional constraint $\chi\le(|\boldsymbol{r}-\boldsymbol{r}'|-1)/(R_0-1)$ is valid up to a time that scales aysmptotically as $|\boldsymbol{r}-\boldsymbol{r}'|^{(\alpha-d)/(d+1)}$, which is already qualitatively larger than the light-cone scaling $|\boldsymbol{r}-\boldsymbol{r}'|^{(\alpha-d)/(\alpha+1)}$. 

\section{Full proof of the clustering properties}

\subsection{Lieb-Robinson bound for short time}
We recall that the main ingredient for proving the clustering properties is nothing but the Lieb-Robinson bound. However, our improved Lieb-Robinson bound only applies for $t>\textcolor{black}{t_{\rm c}}$, while the proof requires a bound valid also for short time scales. Fortunately, $\textcolor{black}{t_{\rm c}}$ is an $\mathcal{O}(1)$ constant and it turns out that the exponentially weaker bound in the seminal paper by Hasting and Koma \cite{Hastings2006} is good enough:
\begin{lemma}[Hastings-Koma bound] For any $\alpha$-decaying Hamiltonian $H$ with $\alpha>d$, we have
\begin{equation}
    \|P_{\boldsymbol{r}'} e^{-iHt} P_{\boldsymbol{r}}\|\le \delta_{\boldsymbol{r}'\boldsymbol{r}}+\frac{e^{\lambda t}-1}{(|\boldsymbol{r}-\boldsymbol{r}'|+1)^\alpha},
\end{equation}
where $\lambda= c_2 J$ with $c_2$ given in Proposition~\ref{Prop:rp} (actually one can take $c_2=2^{\alpha+1}c_1$).
\end{lemma}
\emph{Proof}: Taking $\Theta=0$ in Proposition~\ref{Prop:rp} and using the notations for the original lattice, we find
\begin{equation}
    \sum_{\boldsymbol{r}\in\Lambda}\frac{1}{(|\boldsymbol{r}_1-\boldsymbol{r}|+1)^\alpha}\frac{1}{(|\boldsymbol{r}-\boldsymbol{r}_2|+1)^\alpha}
    \le\frac{c_2}{(|\boldsymbol{r}_1-\boldsymbol{r}_2|+1)^\alpha}.
    \label{b2}
\end{equation}
Applying the above result (\ref{b2}) to the Taylor expansion of $e^{-iHt}$ in the usual Schr\"odinger picture, we obtain
\begin{equation}
\begin{split}
\| P_{\boldsymbol{r}'} e^{-iHt} P_{\boldsymbol{r}}\| &\le \delta_{\boldsymbol{r}'\boldsymbol{r}}+ \sum^\infty_{n=1} \frac{t^n}{n!} \sum_{\{\boldsymbol{r}_j\in\Lambda\}^{n-1}_{j=1}} \|P_{\boldsymbol{r}'} HP_{\boldsymbol{r}_{n-1}} \|\cdots \|P_{\boldsymbol{r}_2} H P_{\boldsymbol{r}_1}\| \|P_{\boldsymbol{r}_1} H P_{\boldsymbol{r}}\| \\
&\le \delta_{\boldsymbol{r}'\boldsymbol{r}} + \sum^\infty_{n=1} \frac{t^n}{n!}\sum_{\{\boldsymbol{r}_j\in\Lambda\}^{n-1}_{j=1}} \frac{J}{(|\boldsymbol{r}' - \boldsymbol{r}_{n-1}|+1)^\alpha} \cdots \frac{J}{(|\boldsymbol{r}_2 - \boldsymbol{r}_1|+1)^\alpha} \frac{J}{(|\boldsymbol{r}_1 - \boldsymbol{r}|+1)^\alpha} \\
&\le \delta_{\boldsymbol{r}'\boldsymbol{r}} + \sum^\infty_{n=1} \frac{t^n}{n!} \frac{J^nc_2^{n-1}}{(|\boldsymbol{r} - \boldsymbol{r}'| + 1)^\alpha} 
= \delta_{\boldsymbol{r}'\boldsymbol{r}} + \frac{e^{c_2 J t} -1}{c_2(|\boldsymbol{r} - \boldsymbol{r}'| +1)^\alpha},
\end{split}
\label{HKb}
\end{equation}
which is the desired result.\hfill$\Box$

\subsection{Covariance matrix of the ground state}
As already mentioned in the main text, the basic idea largely follows that for interacting systems \cite{Hastings2006,Tran2020}. For the convenience of unfamiliar readers, we reproduce the proof below in the language of free fermions. We believe the detailed proof is helpful for understanding not only how our new Lieb-Robinson bound updates the state of the art, but also our new result in the next subsection on Green's functions obtained using a similar strategy. 

\begin{figure}[!t]
\begin{center}
\includegraphics[width=12cm, clip]{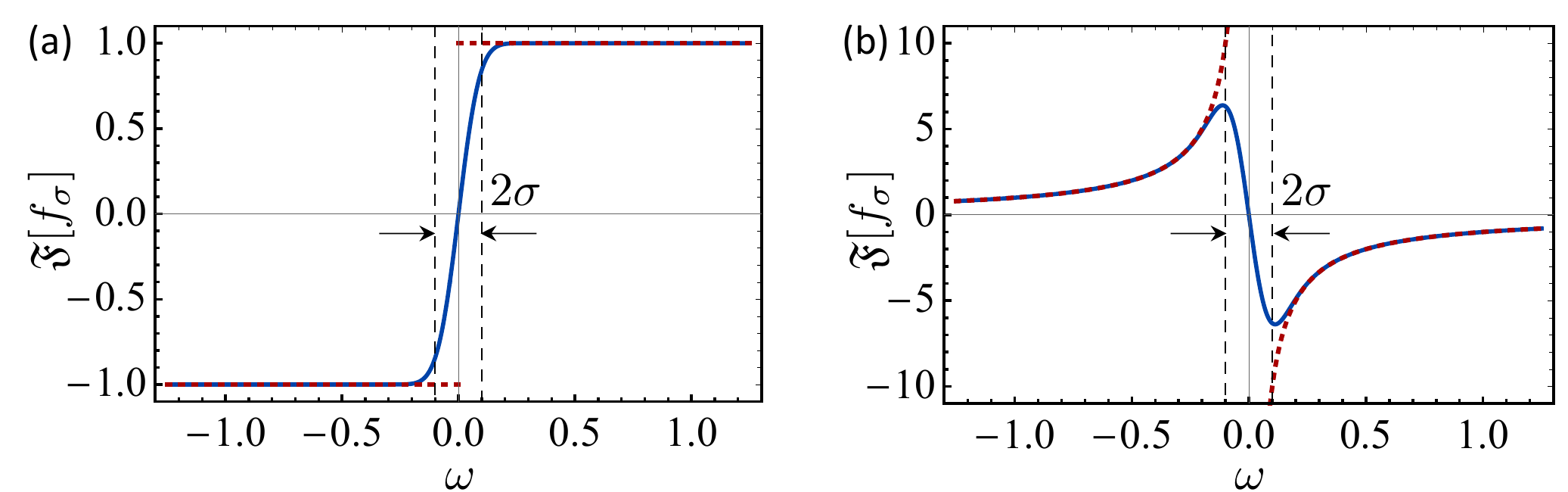}
\end{center}
   \caption{Fourier transforms of the approximate (blue solid curves) and exact (red dotted curves) filters for proving the clustering properties of (a) ground-state covariance matrices and (b) Green's functions (with $z=0$). Here $\sigma=0.1$ in both cases.}
      \label{FT}
\end{figure}

Suppose that $H$ is gapped over $[-\Delta,\Delta]$ with $\Delta > 0$, a perfect filter $f(t)$ whose Fourier transform satisfies
\begin{equation}
\mathfrak{F}[f](\omega)\equiv\int^{\infty}_{-\infty} \frac{dt}{2\pi } f(t) e^{-i\omega t} = {\rm sgn}(\omega),\;\;\;\;\forall |\omega|\ge \Delta
\end{equation}
will yield 
\begin{equation}
\int^{\infty}_{-\infty} \frac{dt}{2\pi } e^{-iHt}f(t) ={\rm sgn}H=\mathbb{1}-2C. 
\end{equation}
For technical reasons that will become clear later, we use an approximate filter $f_\sigma(t)$ that satisfies (see Fig.~\ref{FT}(a))
\begin{equation}
\mathfrak{F}[f_\sigma](\omega)
=  {\rm erf}\left(\frac{\omega}{\sigma}\right),\;\;\;\;
{\rm erf}(z)\equiv \frac{2}{\sqrt{\pi}} \int^z_0 dx e^{-x^2},
\end{equation}
which can be explicitly determined to be 
\begin{equation}
f_\sigma(t) = \frac{2i e^{-\frac{1}{4}\sigma^2t^2}}{t}.
\end{equation}
The error arising from this approximation can be estimated as
\begin{equation}
\left\| \int^{\infty}_{-\infty} \frac{dt}{2\pi } e^{-iHt}f_\sigma(t) - {\rm sgn}H \right\|\le \left|{\rm erf}\left(\frac{\Delta}{\sigma}\right) - 1\right| < \frac{\sigma}{\sqrt{\pi} \Delta} e^{-\frac{\Delta^2}{\sigma^2}}.
\end{equation}
The above result can be further used to bound $\|C_{\boldsymbol{r}\boldsymbol{r}'}\|$ 
as
\begin{equation}
\|C_{\boldsymbol{r}\boldsymbol{r}'}\| < \int^\infty_{-\infty} \frac{dt}{2\pi} \frac{e^{-\frac{1}{4}\sigma^2t^2}}{|t|}B_{\boldsymbol{r}\boldsymbol{r}'}(t) + \frac{\sigma}{2\sqrt{\pi} \Delta} e^{-\frac{\Delta^2}{\sigma^2}}+\frac{1}{2}\delta_{\boldsymbol{r}\boldsymbol{r}'},
\label{Gb}
\end{equation}
where $B_{\boldsymbol{r}\boldsymbol{r}'}(t)$ could be an arbitrary Lieb-Robinson bound: 
\begin{equation}
B_{\boldsymbol{r}\boldsymbol{r}'}(t) \ge \| P_{\boldsymbol{r}} e^{-iHt} P_{\boldsymbol{r}'}\|.
\label{bp}
\end{equation}
Note that in the limit of a perfect filter, i.e., $\sigma\to0$, the time integral in Eq.~(\ref{Gb}) diverges if we use the trivial bound $B_{\boldsymbol{r}\boldsymbol{r}'}(t)=1$ for large $t$. This explains the reason why the filter is chosen to be an approximate one. While it is certainly possible to choose an exact filter whose Fourier transform differs considerably from ${\rm sgn}(\omega)$ on $(-\Delta,\Delta)$, it seems difficult (probably impossible) to make the tail as small as $e^{-\mathcal{O}(t^2)}$ and thus the time integral will blow up, leading to a weaker bound.

We proceed by explicitly evaluating the rhs of Eq.~(\ref{Gb}) using the Lieb-Robinson bound:
\begin{equation}
B_{\boldsymbol{r}\boldsymbol{r}'}(t) =\left\{
\begin{array}{ll} 
\min\{\frac{e^{\lambda|t|}-1}{c_2(|\boldsymbol{r}-\boldsymbol{r}'|+1)^\alpha},1\}, & t\le \max\{\textcolor{black}{t_{\rm c}},0\};   \\
\min\left\{\frac{e^{\lambda|t|}-1}{c_2(|\boldsymbol{r}-\boldsymbol{r}'|+1)^\alpha},\frac{K(|t|)}{(|\boldsymbol{r} - \boldsymbol{r}'|+1)^\alpha}, 1\right\}, & t>\max\{\textcolor{black}{t_{\rm c}},0\}.
\end{array}
\right.
\label{Bt}
\end{equation}
Note that here $t$ is allowed to be negative. Such a generalization can be easily achieved via $H\to -H$, which does not alter any parameters in the bounds. 
We introduce two time parameters $\tau_1$, $\tau_2$ with $\tau_2\ge\tau_1>\textcolor{black}{t_{\rm c}}$ and apply the tightened Lieb-Robinson bound for the intermediate time interval $t\in[\tau_1,\tau_2]$, obtaining
\begin{equation}
\begin{split}
\|C_{\boldsymbol{r}\boldsymbol{r}'}\| &< \int^{\tau_1}_0 \frac{dt}{\pi}\frac{e^{\lambda t}-1}{c_2t(|\boldsymbol{r}-\boldsymbol{r}'|+1)^\alpha} +
\int^{\tau_2}_{\tau_1}\frac{dt}{\pi t} \frac{K(t)}{(|\boldsymbol{r} - \boldsymbol{r}'|+1)^\alpha}
+  \int^\infty_{\tau_2} \frac{dt}{\pi} \frac{e^{-\frac{1}{4}\sigma^2t^2}}{t} 
+ \frac{\sigma}{2\sqrt{\pi} \Delta} e^{-\frac{\Delta^2}{\sigma^2}} + \frac{1}{2}\delta_{\boldsymbol{r}\boldsymbol{r}'}\\
&<\frac{e^{\lambda\tau_1}}{\pi c_2 (|\boldsymbol{r} - \boldsymbol{r}'| +1)^\alpha} + \frac{\tau_2 K (\tau_2)}{\pi\tau_1 (|\boldsymbol{r} - \boldsymbol{r}'| +1)^\alpha} + \frac{2e^{-\frac{1}{4}\sigma^2\tau_2^2}}{\pi\sigma^2\tau_2^2} + \frac{\sigma}{2\sqrt{\pi}\Delta} e^{-\frac{\Delta^2}{\sigma^2}} + \frac{1}{2}\delta_{\boldsymbol{r}\boldsymbol{r}'},
\end{split}
\label{Crrll}
\end{equation}
where we have used the fact that $K(t)$ increases monotonically with $t$. To suppress the error from the approximate filter as proportional to  $(|\boldsymbol{r}-\boldsymbol{r}'|+1)^{-\alpha}$, we choose $\sigma$ and $\tau_2$ such that
\begin{equation}
\frac{1}{(|\boldsymbol{r} - \boldsymbol{r}'|+1)^\alpha} = e^{-\frac{1}{4}\sigma^2\tau_2^2} = e^{-\frac{\Delta^2}{\sigma^2}}\;\;\;\;\Rightarrow\;\;\;\;
\sigma^2=\frac{2\Delta}{\tau_2},\;\;\tau_2=\frac{2\alpha}{\Delta}\log(|\boldsymbol{r}-\boldsymbol{r}'|+1).
\end{equation}
While this is probably not optimal (even qualitatively), we choose $\tau_1 = (\alpha -1)/v  > \max\{(\alpha - R_0 -1)/v,0\}$ and end up with
\begin{equation}
\|C_{\boldsymbol{r}\boldsymbol{r}'}\| \le \frac{P(\frac{2\alpha}{\Delta}\log(|\boldsymbol{r} - \boldsymbol{r}'| + 1))}{(|\boldsymbol{r} - \boldsymbol{r}'| + 1)^\alpha},
\label{Crrp}
\end{equation}
where $P(t)$ scales at most polynomially with $t$:
\begin{equation}
P(t)=
\frac{vtK(t)}{\pi(\alpha -1)} + \frac{e^{(\alpha-1)\frac{\lambda}{v}}}{\pi c_2} + \frac{v}{\pi(\alpha-1)\Delta} + \sqrt{\frac{v}{2\pi(\alpha-1)\Delta}}+\frac{1}{2}. 
\end{equation}
\textcolor{black}{We note that while we assumed $\tau_2\ge\tau_1$ in the derivation of Eq.~(\ref{Crrp}), the result itself holds true regardless of this assumption since the second line of Eq.~(\ref{Crrll}) is always an overestimation. Importantly,} 
this result implies that, for arbitrarily small $\epsilon>0$, there exist a constant $\textcolor{black}{K}_\epsilon$ such that $\|\textcolor{black}{C_{\boldsymbol{r}\boldsymbol{r}'}}\| \le \textcolor{black}{K}_\epsilon/( |\boldsymbol{r} - \boldsymbol{r}'| + 1)^{\alpha-\epsilon}$.

\subsection{Green's function}
We move on to the Green's function:
\begin{equation}
G_{\boldsymbol{r}\boldsymbol{r}'}(z) = P_{\boldsymbol{r}} \frac{1}{z-H} P_{\boldsymbol{r}'},
\label{GFapp}
\end{equation}
which is an $|I|\times |I|$ matrix. 
Suppose there is a filter $f(t)$ satisfying
\begin{equation}
\mathfrak{F}[f](\omega) 
= \frac{1}{z-\omega},
\end{equation}
we have 
\begin{equation}
G_{\boldsymbol{r}\boldsymbol{r}'}(z) = \int^\infty_{-\infty} \frac{dt}{2\pi} f(t)P_{\boldsymbol{r}} e^{-iHt} P_{\boldsymbol{r}'}.
\end{equation}
However, similar to the case of covariance matrices, such an exact filter is ill-defined and not helpful for deriving the clustering properties. Instead, we consider an approximate filter $f_\sigma(t)$ with a tunable positive parameter $\sigma$ (see Fig.~\ref{FT}(b)):
\begin{equation}
\mathfrak{F}[f_\sigma](\omega) = \frac{1-e^{-(\omega-z)(\omega-\bar z)/\sigma^2}}{z-\omega},
\label{afGF}
\end{equation}
where the rhs is free from the pole singularity at $\omega=z$ but remains a good approximation of $(z-\omega)^{-1}$ if $\omega$ is far from $z$ or/and $\sigma$ is small enough. One can check 
the following explicit expression and properties of $f_\sigma(t)$:
\begin{proposition}[approximate filter for the Green's function] The filter $f_\sigma(t)$ satisfying Eq.~(\ref{afGF}) is given by
\begin{equation}
 f_\sigma(t)=i\pi e^{izt}\left[{\rm erf}\left(\frac{\sigma }{2}t +\frac{\gamma}{\sigma}\right) - {\rm sgn}(t) \right],\;\;\;\;\gamma=\frac{i}{2}(z-\bar z)=-{\rm Im}z,
 \label{fst}
\end{equation}
whose absolute value increases (decreases) monotonically on $(-\infty,0)$ ($(0,\infty)$). 
\end{proposition}
\emph{Proof}: 
After an appropriate shift and rescaling, it suffices to show that $\forall a\in\mathbb{R}$
\begin{equation}
\frac{i}{2}\int^\infty_{-\infty} dt \left[{\rm erf}\left(\frac{t }{2} - a\right) - {\rm sgn}(t) \right]e^{-i\omega t}
= -\frac{1-e^{-\omega(\omega + 2ia)}}{\omega}.
\end{equation}
Noting that the integrand vanishes (super-exponentially) when $t\to\pm\infty$, we can perform an integral by parts, obtaining
\begin{equation}
\frac{1}{2\omega}\int^\infty_{-\infty} dt \left[ \frac{1}{\sqrt{\pi}}e^{-(\frac{t}{2}-a)^2} - 2\delta(t) \right]e^{-i\omega t}
= - \frac{1}{\omega} + \frac{e^{-\omega(\omega+2ia)}}{2\sqrt{\pi}\omega} \int^\infty_{-\infty} dt e^{-(\frac{t}{2}-a+i\omega)^2} = - \frac{1}{\omega} + \frac{e^{-\omega(\omega+2ia)}}{\omega},
\end{equation}
which gives the desired result. Regarding the absolute value of Eq.~(\ref{fst}), which reads
\begin{equation}
|f_\sigma(t)| 
= \pi e^{\gamma t} \left[1- {\rm sgn}(t) {\rm erf}\left(\frac{\sigma }{2}t + \frac{\gamma}{\sigma}\right)\right],
\end{equation}
we first show that it 
decreases monotonically on $t\in(0,\infty)$. For $t>0$, the derivative is given by
\begin{equation}
\frac{d}{dt}|f_\sigma(t)| = \pi\gamma e^{\gamma t}\left[1- {\rm erf}\left(\frac{\sigma }{2}t + \frac{\gamma}{\sigma}\right)\right] - \sqrt{\pi}\sigma e^{\gamma t} e^{-(\frac{\sigma }{2}t + \frac{\gamma}{\sigma})^2}.
\label{dp}
\end{equation}
If $\gamma\le0$, the derivative (\ref{dp}) is obviously negative. Otherwise, $\gamma>0$, we have
\begin{equation}
1- {\rm erf}\left(\frac{\sigma }{2}t + \frac{\gamma}{\sigma}\right) = 
\frac{2}{\sqrt{\pi}} \int^\infty_{\frac{\sigma }{2}t + \frac{\gamma}{\sigma}} dx e^{-x^2}
< \frac{2}{\sqrt{\pi}(\frac{\sigma }{2}t + \frac{\gamma}{\sigma})} \int^\infty_{\frac{\sigma }{2}t + \frac{\gamma}{\sigma}} dx x e^{-x^2}
=  \frac{e^{-(\frac{\sigma }{2}t + \frac{\gamma}{\sigma})^2}}{\sqrt{\pi}(\frac{\sigma }{2}t + \frac{\gamma}{\sigma})},
\end{equation}
leading to
\begin{equation}
\frac{d}{dt}|f_\sigma(t)| < - \frac{\sqrt{\pi}\sigma^3 t}{\sigma^2 t + 2\gamma} e^{-(\frac{\sigma }{2}t + \frac{\gamma}{\sigma})^2+\gamma t} < 0.
\end{equation}
We then show that $|f_\sigma(t)|$ increases monotonically on $t\in(-\infty,0)$. For $t<0$, the derivative is given by
\begin{equation}
\frac{d}{dt}|f_\sigma(t)| = \pi\gamma e^{\gamma t}\left[1+ {\rm erf}\left(\frac{\sigma }{2}t + \frac{\gamma}{\sigma}\right)\right] + \sqrt{\pi}\sigma e^{\gamma t} e^{-(\frac{\sigma }{2}t + \frac{\gamma}{\sigma})^2}.
\label{dm}
\end{equation}
If $\gamma\ge0$, the derivative (\ref{dm}) is obviously positive. Otherwise, $\gamma<0$, following a procedure similar to the previous case, we can show that 
\begin{equation}
\frac{d}{dt}|f_\sigma(t)| > \frac{\sqrt{\pi}\sigma^3 t}{\sigma^2 t + 2\gamma} e^{-(\frac{\sigma }{2}t + \frac{\gamma}{\sigma})^2+\gamma t} > 0.
\end{equation}
In particular, these results imply $|f_\sigma(t)|\le|f_\sigma(0^+)|=\pi [1-{\rm erf}(\gamma/\sigma)]$ $\forall t>0$ and $|f_\sigma(t)|\le|f_\sigma(0^-)|=\pi [1+{\rm erf}(\gamma/\sigma)]$ $\forall t<0$. \hfill$\Box$ \\
In addition, one can easily confirm that this approximate filter (\ref{fst}) vanishes rapidly as $e^{-\mathcal{O}(t^2)}$ for large $t$

We proceed in a quite parallel way as the previous subsection. Suppose $B_{\boldsymbol{r}\boldsymbol{r}'}(t)$ is a Lieb-Robinson bound (\ref{bp}), 
we know that the norm of the Green's function (\ref{GFapp}) admits the following upper bound:
\begin{equation}
\|G_{\boldsymbol{r}\boldsymbol{r}'}(z) \| \le \int^\infty_{-\infty}\frac{dt}{2\pi}|f_\sigma(t)|B_{\boldsymbol{r}\boldsymbol{r}'}(t) +  \frac{e^{-\Delta(z)^2/\sigma^2}}{\Delta(z)},
\label{GzB}
\end{equation}
where $\Delta(z)\equiv\|(z-H)^{-1}\|^{-1}=\min\{|z-\lambda|: \lambda\in{\rm Spec}(H)\}$ denotes the distance between $z$ and the spectrum of $H$. Recalling that there is always a trivial bound $B_{\boldsymbol{r}\boldsymbol{r}'}(t)\le 1$ and $B_{\boldsymbol{r}\boldsymbol{r}'}(-t)=B_{\boldsymbol{r}\boldsymbol{r}'}(t)$, we can introduce a time parameter $\tau > 2|\gamma|/\sigma^2$ 
to further bound Eq.~(\ref{GzB}) by
\begin{equation}
\begin{split}
\|G_{\boldsymbol{r}\boldsymbol{r}'}(z) \| &\le \int^\tau_{-\tau} \frac{dt}{2\pi} |f_\sigma(t)| B_{\boldsymbol{r}\boldsymbol{r}'}(t) 
+ \left(\int^\infty_\tau + \int^{-\tau}_{-\infty}\right) \frac{dt}{2\pi} |f_\sigma(t)| + \frac{e^{-\Delta(z)^2/\sigma^2}}{\Delta(z)} \\
&<  \int^\tau_0 dt B_{\boldsymbol{r}\boldsymbol{r}'}(t) + \frac{1}{\sqrt{\pi}\sigma} \frac{e^{-\frac{1}{4}\sigma^2\tau^2 - \frac{\gamma^2}{\sigma^2}}}{\frac{1}{4}\sigma^2\tau^2 - \frac{\gamma^2}{\sigma^2}} + \frac{e^{-\Delta(z)^2/\sigma^2}}{\Delta(z)} 
\end{split}
\end{equation}
where we have use the monotonicity of $|f_\sigma(t)|$, $|f_\sigma(0^+)| + |f_\sigma(-0^+)|=2\pi$ and
\begin{equation}
\begin{split}
|f_\sigma(t)|= 2\sqrt{\pi} e^{\gamma t} \int^\infty_{\frac{\sigma}{2}t + \frac{\gamma}{\sigma}} dx e^{-x^2}
< \frac{\sqrt{\pi} e^{\gamma t}}{\frac{\sigma}{2}t + \frac{\gamma}{\sigma}} \int^\infty_{(\frac{\sigma}{2}t + \frac{\gamma}{\sigma})^2} dy e^{-y} 
= \frac{\sqrt{\pi}}{\frac{\sigma}{2}t + \frac{\gamma}{\sigma}} e^{-\frac{\sigma^2}{4}t^2 - \frac{\gamma^2}{\sigma^2}},\;\;\;\;{\rm if}\; t>\max\left\{-\frac{2\gamma}{\sigma^2},0\right\}; \\
|f_\sigma(t)|= 2\sqrt{\pi} e^{\gamma t} \int^{\frac{\sigma}{2}t + \frac{\gamma}{\sigma}}_{-\infty} dx e^{-x^2}
< -\frac{\sqrt{\pi} e^{\gamma t}}{\frac{\sigma}{2}t + \frac{\gamma}{\sigma}} \int^\infty_{(\frac{\sigma}{2}t + \frac{\gamma}{\sigma})^2} dy e^{-y} 
=- \frac{\sqrt{\pi}}{\frac{\sigma}{2}t + \frac{\gamma}{\sigma}} e^{-\frac{\sigma^2}{4}t^2 - \frac{\gamma^2}{\sigma^2}},\;\;\;\;{\rm if}\; t<\min\left\{-\frac{2\gamma}{\sigma^2},0\right\}.
\end{split}
\end{equation}
Recalling that $B_{\boldsymbol{r}\boldsymbol{r}'}(t)$ is given in Eq.~(\ref{Bt}), which can actually be simplified into $B_{\boldsymbol{r}\boldsymbol{r}'}(t) = p(t)/(|\boldsymbol{r}-\boldsymbol{r}'|+1)^\alpha$ with $p(t)$ being at most polynomially large in $t$ (note that $e^{\lambda t}-1$ up to a finite time $\textcolor{black}{t_{\rm c}}$ can be upper bounded by, e.g., $(e^{\lambda \textcolor{black}{t_{\rm c}}}-1)t/\textcolor{black}{t_{\rm c}}$), one can choose
\begin{equation}
\sigma= \frac{\Delta (z)}{\sqrt{\alpha \log(|\boldsymbol{r} - \boldsymbol{r}'|+1)+1}},\;\;\;\;
\tau= \frac{2\sqrt{\Delta(z)^2 + \gamma^2}}{\sigma^2} =\frac{2\sqrt{\Delta(z)^2 + \gamma^2}}{\Delta(z)^2}[\alpha\log(|\boldsymbol{r} - \boldsymbol{r}'|+1)+1] > \frac{2|\gamma|}{\sigma^2},
\end{equation} 
such that 
\begin{equation}
\begin{split}
\|G_{\boldsymbol{r}\boldsymbol{r}'}(z) \| &\le \frac{P(2\sqrt{\Delta(z)^2 + \gamma^2}[\alpha\log(|\boldsymbol{r} - \boldsymbol{r}'|+1)+1]/\Delta(z)^2)}{(|\boldsymbol{r} - \boldsymbol{r}'|+1)^\alpha} \\
&+\frac{1}{\Delta(z)\sqrt{\pi[\alpha\log(|\boldsymbol{r} - \boldsymbol{r}'| +1)+1]}[e(|\boldsymbol{r} - \boldsymbol{r}'|+1)^\alpha]^{1+\frac{2\gamma^2}{\Delta(z)^2}}}
+ \frac{1}{e\Delta(z)(|\boldsymbol{r} - \boldsymbol{r}'|+1)^\alpha},
\end{split}
\label{Grrp}
\end{equation}
where $P(t)\equiv \int^t_{0} dt' p(t')$ is again polynomially large in $t$ so we have obtained the desired result. 

\subsection{Example of bound states}
\begin{figure}[!t]
\begin{center}
\includegraphics[width=12cm, clip]{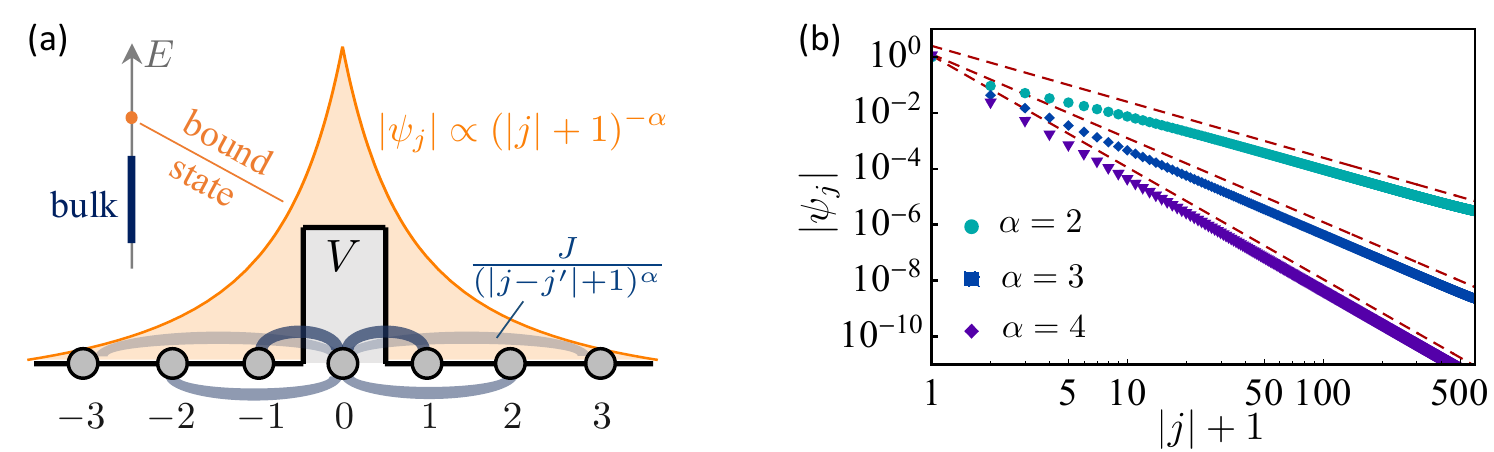}
\end{center}
   \caption{(a) 1D lattice with $\alpha$-decaying hopping amplitudes and an on-site impurity potential $V$ at the origin, as described by Eq.~(\ref{HJV}). The orange region shows the profile of the bound state. (b) Numerical results for the bound-state profiles at different $\alpha$. The red dashed lines indicates exact $\alpha$-decaying. Here $V=3J$ and (system size) $L=2000$.}
      \label{BS}
\end{figure}

As mentioned in the main text, the clustering property of the Green's function applies straightforwardly to the bound states induced by local impurities. The arguably simplest example could be a single-band 1D chain with long-range hopping and an on-site impurity potential:
\begin{equation}
    \hat H = \sum_{j,j'}\frac{J}{(|j-j'|+1)^\alpha}(\hat c^\dag_j \hat c_{j'} + {\rm H.c.}) + V\hat c_0^\dag \hat c_0.
    \label{HJV}
\end{equation}
See Fig.~\ref{BS}(a) for a schematic illustration. In practical calculations for a finite chain with $L$ sites under the periodic boundary condition, we replace the hopping amplitude $J/(|j-j'|+1)^\alpha$ by $J[1/(|j-j'|+1)^\alpha + 1/(L-|j-j'|+1)^\alpha]$. As shown in Fig.~\ref{BS}(b), the tail of the bound state indeed follows the same algebraic decay as the Hamiltonian.

\section{Some results on long-range topological phases}

\subsection{Connectivity to short-range phases}
We recall our protocol of deforming an $\alpha$-decaying ($\alpha>d$) Hamiltonian $H$ into a short-range one is simply multiplying an factor $e^{-\kappa|\boldsymbol{r}-\boldsymbol{r}'|}$ to each block $H_{\boldsymbol{r}\boldsymbol{r}'}$. We first show that the family of Hamiltonians generated by varying $\kappa$ is at least H\"older continuous. To see this, we only have to upper bound the norm difference between two deformed Hamiltonians with $\kappa\neq\kappa'$ by  
\begin{equation}
\begin{split}
    \|H_\kappa-H_{\kappa'}\|&\le\max_{\boldsymbol{r}}\sum_{\boldsymbol{r}'}\|H_{\kappa,\boldsymbol{r}\boldsymbol{r}'}-H_{\kappa',\boldsymbol{r}\boldsymbol{r}'}\|=\max_{\boldsymbol{r}}\sum_{\boldsymbol{r}'}\left|e^{-\kappa|\boldsymbol{r}-\boldsymbol{r}'|}-e^{-\kappa'|\boldsymbol{r}-\boldsymbol{r}'|}\right|\|H_{\boldsymbol{r}\boldsymbol{r}'}\|\\
    &\le\max_{\boldsymbol{r}}\sum_{\boldsymbol{r}'}\left|1-e^{-|\kappa-\kappa'||\boldsymbol{r}-\boldsymbol{r}'|}\right|\|H_{\boldsymbol{r}\boldsymbol{r}'}\|.
\end{split}
\end{equation}
To proceed, we separate the above sum into to parts:
\begin{equation}
    \sum_{\boldsymbol{r}':|\boldsymbol{r}'-\boldsymbol{r}|<\xi}|\kappa-\kappa'||\boldsymbol{r}-\boldsymbol{r}'|\|H_{\boldsymbol{r}\boldsymbol{r}'}\|+
    \sum_{\boldsymbol{r}':|\boldsymbol{r}'-\boldsymbol{r}|\ge\xi}\|H_{\boldsymbol{r}\boldsymbol{r}'}\|
    \le c_1J\xi|\kappa-\kappa'|+\frac{c_1J}{(\xi+1)^{\alpha-d}},
\end{equation}
where $1-e^{-x}\le x$ $\forall x\ge0$ and Eq.~(\ref{b1}) have been used, and $\xi$ is a parameter to be determined. Choosing $\xi=\lfloor|\kappa-\kappa'|^{-1/(\alpha-d+1)}\rfloor$, we obtain
\begin{equation}
    \|H_\kappa-H_{\kappa'}\|\le 2c_1J|\kappa-\kappa'|^{\frac{\alpha-d}{\alpha-d+1}},
    \label{HkHk}
\end{equation}
which is the desired H\"older continuity property.

In particular, suppose that the original Hamiltonian $H_0=H$ is gapped over $[-\Delta,\Delta]$, according to Weyl's perturbation theorem \cite{Bhatia1997}, we know that $H_\kappa$ is also gapped as long as $\|H_\kappa - H_0\|< \Delta$. Taking $\kappa'=0$ in Eq.~(\ref{HkHk}), we conclude that this condition can be ensured by
\begin{equation}
    \kappa< \left(\frac{\Delta}{2c_1J}\right)^{\frac{\alpha-d+1}{\alpha-d}}.
    \label{kd}
\end{equation}
Note that by definition $\Delta\le\|H\|\le c_1J$, so Eq.~(\ref{kd}) means $\kappa$ should be sufficiently small, especially for $\alpha$ close to $d$.

\subsection{Gapped nature of the deformation path}
We show that the continuous path connecting the trivial and topological short-range phases: 
\begin{equation}
 h_\lambda(\boldsymbol{k})=\left\{
 \begin{array}{ll} 
 (1-2\lambda)h_0(\boldsymbol{k})+2\lambda h_{\rm fD}(\boldsymbol{k}), & \lambda\in[0,0.5];  \\
 2(1-\lambda)h_{\rm fD}(\boldsymbol{k})+(2\lambda-1)h_{\rm Topo}(\boldsymbol{k}), & \lambda\in(0.5,1],
 \end{array}
 \right.
 \label{hl}
\end{equation}
is gapped for any $\lambda$. To this end, it suffices to show $h_\lambda(\boldsymbol{k})^2>0$. For $\lambda\in[0,0.5]$, noting that $\{h_0(\boldsymbol{k}),h_{\rm fD}(\boldsymbol{k})\}=0$ and $h_0(\boldsymbol{k})^2=h_{\rm fD}(\boldsymbol{k})^2=1$, we have
\begin{equation}
   h_\lambda(\boldsymbol{k})^2=(1-2\lambda)^2+4\lambda^2\ge\frac{1}{2}>0.
   \label{hl2f}
\end{equation}
For $\lambda\in[0.5,1]$, after some straightforward calculations, we obtain
\begin{equation}
\begin{split}
    h_\lambda(\boldsymbol{k})^2&=4(1-\lambda)^2+4(1-\lambda)(2\lambda-1)\sqrt{\sum^d_{\mu=1}\sin^2k_\mu}+(2\lambda-1)^2\left(\sum^d_{\mu=1}\sin^2k_\mu\right)+(2\lambda-1)^2\left(\sum^d_{\mu=1}\cos k_\mu -d +1\right)^2 \\
    &\ge 4(1-\lambda)^2+(2\lambda-1)^2\left[2\sum_{1\le\mu<\nu\le d}\cos k_\mu\cos k_\nu-2(d-1)\left(\sum^d_{\mu=1}\cos k_\mu\right)+d^2-d+1\right] \\
    &=4(1-\lambda)^2+(2\lambda-1)^2\left[2\sum_{1\le\mu<\nu\le d}(1-\cos k_\mu)(1-\cos k_\nu)+1\right] \\
    &\ge4(1-\lambda)^2+(2\lambda-1)^2\ge\frac{1}{2}>0.
\end{split}
\label{hl2s}
\end{equation}
Combining Eqs.~(\ref{hl2f}) and (\ref{hl2s}), we have confirmed that Eq.~(\ref{hl}) is indeed a path of gapped Hamiltonians.

\subsection{Bound on real-space decay for Bloch Hamiltonians with singular points} 
In the main text, we have numerically demonstrated 
that a continuous interpolation between trivial and topological $d$D free-fermion phases is possible if the ground-state covariance matrix is allowed to be $\alpha$-decaying with $\alpha<d$. 
Here, we provide further analytical argument on this observation, making it clear that the algebraic decay in real space arises from the momentum-space singularity. 

\begin{figure}[!t]
\begin{center}
\begin{tikzpicture}[scale=0.5]
\draw[fill=blue!2.5!white] (-4,-4) rectangle (4,4);
\draw[fill=red!10!white] (-4,-1) rectangle (4,1);
\draw[fill=blue!20!white] (0,0) circle (1);
\begin{scope}[>=latex] 
\draw[->] (-5,0) -- (5,0);
\draw[->] (0,-5) -- (0,5);
\end{scope}
\draw[ultra thick] (0,-1) -- (0,1);
\Text[x=5.4,y=0,fontsize=\small]{$k_\mu$}
\Text[x=0,y=5.4,fontsize=\small]{$k_\perp$}
\Text[x=1.3,y=0.7,fontsize=\small,color=blue!50!black]{$\Omega^d_\epsilon$}
\Text[x=0,y=-1.25,fontsize=\tiny]{$\Omega^{d-1}_\epsilon$}
\Text[x=-0.3,y=-0.3,fontsize=\small]{$O$}
\Text[x=3.8,y=-0.3,fontsize=\small]{$\pi$}
\Text[x=-0.2,y=3.7,fontsize=\small]{$\pi$}
\Text[x=-4,y=-0.3,fontsize=\small]{$-\pi$}
\Text[x=0,y=-4.3,fontsize=\small]{$-\pi$}
\Text[x=1.2,y=-0.25,fontsize=\small]{$\epsilon$}
\draw[fill=black] (1,0) circle (0.05);
\draw[fill=black] (0,0) circle (0.08);
\draw[fill=black] (4,0) circle (0.05) (-4,0) circle (0.05) (0,-4) circle (0.05) (0,4) circle (0.05);
\end{tikzpicture}
\caption{Decomposition of the integral region. The blue circle and thick intervals refer to $\Omega^d_\epsilon$ and $\Omega^{d-1}_\epsilon$, respectively. The middle term in Eq.~(\ref{Crb}) is associated with the boundaries between blue circle and red regions. The figure should be understood as a 2D slice spanned by $k_\mu$ and a specific direction $k_\perp$ perpendicular to $k_\mu$.}
      \label{ID}
      \end{center}
\end{figure}
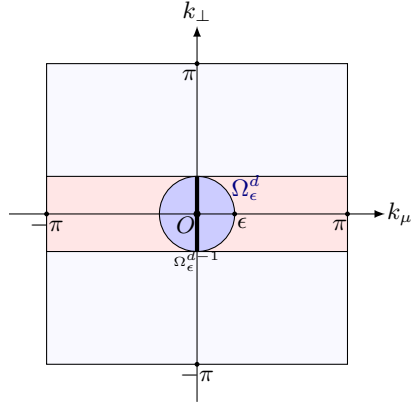

Our argument is based on the following proposition, which generalizes a result in Ref.~\cite{Wahl2014} (corresponding to $q=3$, $d=2$):
\begin{proposition}[Real-space algebraic decay constrained by momentum-space singularity]
For a matrix-value function $M(\boldsymbol{k})$ defined on a $d$D Brillouin zone, suppose that there exist $\mathcal{O}(1)$ constants $b_0,b_1$ and integer $q\ge d$
such that $\forall \mu=1,2,...,d$
\begin{equation}
\left\| \frac{\partial^q M(\boldsymbol{k})}{\partial k_\mu^q} \right\| \le \frac{b_0}{|\boldsymbol{k}|^d},\;\;\;\;
\left\| \frac{\partial^{q-1} M(\boldsymbol{k})}{\partial k_\mu^{q-1}} \right\| \le \frac{b_1}{|\boldsymbol{k}|^{d-1}},
\label{qCd}
\end{equation}
its inverse Fourier transform $M_{\boldsymbol{r}} = \int_{\rm B.Z.} \frac{d^d\boldsymbol{k}}{(2\pi)^d} M(\boldsymbol{k}) e^{i\boldsymbol{k}\cdot \boldsymbol{r}}$
satisfies the $q$-decaying property up to logarithmic corrections.
\end{proposition}
\emph{Proof}: 
For technical reasons, we temporarily adopt the convention $|\boldsymbol{r}|^{-\alpha}$ (rather than $(|\boldsymbol{r}|+1)^{-\alpha}$) for $\alpha$-decay. If we further assume $M(\boldsymbol{k})$ to be bounded, which is typically the case in practice, we know that $M_{\boldsymbol{0}}=\int_{\rm B.Z.}\frac{d^d\boldsymbol{k}}{(2\pi)^d} M(\boldsymbol{k})$ is finite and there is indeed no difference between the two conventions. Similar to Ref.~\cite{Wahl2014}, our main technique is (repeated) integration by parts:
\begin{equation}
M_{\boldsymbol{r}} = \int_{\rm B.Z.} \frac{d^d\boldsymbol{k}}{(2\pi)^d} M(\boldsymbol{k}) e^{i\boldsymbol{k}\cdot \boldsymbol{r}}
 = \left(-\frac{1}{i r_\mu}\right)^p  \int_{\rm B.Z.} \frac{d^d\boldsymbol{k}}{(2\pi)^d}  \frac{\partial^p M(\boldsymbol{k})}{\partial k_\mu^p} e^{i\boldsymbol{k}\cdot \boldsymbol{r}}.
 \label{Gr}
\end{equation}
Here $r_{\mu}$ is chosen to be the largest (in the sense of absolute value) component such that $r_\mu\ge |\boldsymbol{r}|/\sqrt{d}>0$ whenever $\boldsymbol{r}\neq\boldsymbol{0}$. We estimate the rightmost integral in Eq.~(\ref{Gr}) with $p$ chosen to be $q-1$ in two parts, one within $\Omega^d_\epsilon$, the $d$D sphere with radius $\epsilon$ centered at $\boldsymbol{k}=\boldsymbol{0}$, and the other outside (see Fig.~\ref{ID}). For the latter part, we further perform an integration by parts, leading to
\begin{equation}
\begin{split}
\| M_{\boldsymbol{r}}  \| &\le \frac{1}{|r_\mu|^{q-1}} \int_{\Omega_\epsilon} \frac{d^d\boldsymbol{k}}{(2\pi)^d} \frac{b_1}{|\boldsymbol{k}|^{d-1}}  + \frac{1}{|r_\mu|^{q-1}} \left\| \int_{{\rm B.Z.}\backslash \Omega^d_\epsilon} \frac{d^d\boldsymbol{k}}{(2\pi)^d} \frac{\partial^{q-1} M(\boldsymbol{k})}{\partial k_\mu^{q-1}}e^{i\boldsymbol{k}\cdot\boldsymbol{r}} \right\| \\
&\le \frac{b_1d}{(2\sqrt{\pi})^d\Gamma(\frac{d}{2}+1)}\frac{\epsilon}{|r_\mu|^{q-1}} 
+ \frac{1}{|r_\mu|^q} \left\| \int_{\Omega^{d-1}_\epsilon} \frac{d^{d-1}\boldsymbol{k}_{\perp}}{(2\pi)^d} \left.\frac{\partial^{q-1} M(\boldsymbol{k})}{\partial k_\mu^{q-1}}e^{i\boldsymbol{k}\cdot\boldsymbol{r}}\right|^{k_\mu = \sqrt{\epsilon^2-|\boldsymbol{k}_{\perp}|^2}}_{k_\mu = -\sqrt{\epsilon^2-|\boldsymbol{k}_{\perp}|^2}} \right\|
+ \frac{b_0}{|r_\mu|^q} \int_{{\rm B.Z.}\backslash \Omega^d_\epsilon} \frac{d^d\boldsymbol{k}}{(2\pi)^d} \frac{1}{|\boldsymbol{k}|^d} \\
&< \frac{b_1d}{(2\sqrt{\pi})^d\Gamma(\frac{d}{2}+1)}\frac{\epsilon}{|r_\mu|^{q-1}} 
+ \frac{4b_1}{(2\sqrt{\pi})^{d+1}\Gamma(\frac{d+1}{2})}\frac{1}{|r_\mu|^q}
+ \frac{b_0d}{(2\sqrt{\pi})^d\Gamma(\frac{d}{2}+1)}\frac{\log(\sqrt{d}\pi/\epsilon)}{|r_\mu|^q}
\end{split}
\label{Crb}
\end{equation}
where $\Gamma(z)=\int^\infty_0 dy y^{z-1}e^{-y}$ is the Gamma function, $\boldsymbol{k}_{\perp}$ denotes the $d-1$ components other than $k_\mu$ and the middle term arises from the boundary contribution in the integration by parts (do not cancel each other only when $|\boldsymbol{k}_\perp|<\epsilon$, i.e., $\boldsymbol{k}_\perp$ is within $\Omega^{d-1}_\epsilon$). Choosing 
$\epsilon =|r_\mu|^{-1}\ge |\boldsymbol{r}|^{-1}$ in Eq.~(\ref{Crb}) yields
\begin{equation}
\| M_{\boldsymbol{r}}  \| < \frac{c + c'\log|\boldsymbol{r}|}{|\boldsymbol{r}|^q}, 
\end{equation}
where $c$ and $c'$ are two $\boldsymbol{r}$-independent constants:
\begin{equation}
c=\frac{[b_1 + b_0\log(\sqrt{d}\pi)]d^{1+\frac{d}{2}}}{(2\sqrt{\pi})^d \Gamma(\frac{d}{2}+1)}
+ \frac{4b_1d^{\frac{d}{2}}}{(2\sqrt{\pi})^{d+1} \Gamma(\frac{d+1}{2})},\;\;\;\;
c'=\frac{b_0d^{1+\frac{d}{2}}}{(2\sqrt{\pi})^d \Gamma(\frac{d}{2}+1)}.
\end{equation}
So far we have obtained the desired result.  \hfill$\Box$

The above result can be naturally extended to multiple singular points. That is, the condition (\ref{qCd}) can be replaced by:
\begin{equation}
\left\| \frac{\partial^q M(\boldsymbol{k})}{\partial k_\mu^q} \right\| \le \sum^m_{j=1} \frac{b^{(j)}_0}{|\boldsymbol{k}-\boldsymbol{k}^{(j)}|^d},\;\;\;\;
\left\| \frac{\partial^{q-1} M(\boldsymbol{k})}{\partial k_\mu^{q-1}} \right\| \le \sum^m_{j=1} \frac{b^{(j)}_1}{|\boldsymbol{k}-\boldsymbol{k}^{(j)}|^{d-1}}.
\end{equation}
For the $j$th term, we only have to employ the above partial integral technique near $\boldsymbol{k}^{(j)}$. Now returning to our interpolation protocol based on the flattened Dirac Hamiltonian, one may argue that this corresponds to the case of $q=d$. This may be understood by considering the linearized flattened Dirac Hamiltonian
\begin{equation}
    h_{\rm fD}(\boldsymbol{k})\simeq \sum^d_{\mu=1} \frac{k_\mu}{|\boldsymbol{k}|}\Gamma_\mu,
\end{equation}
for which one roughly obtains something like $k_\mu^{-p}$ after $p$ times of partial derivatives with respect to $k_\mu$. The ``degree of singularity" certainly does not change upon linear combination with other Hamiltonians. Moreover, we expect the same ``degree of singularity" for the flattened interpolated Hamiltonians, provided they are gapped. Therefore, not only the interpolated Hamiltonians but also their grond-state covariance matrices should be essentially $d$-decaying.

\end{document}